\documentclass[11pt,english]{article}
\pdfoutput=1
\usepackage{jheppub}
\usepackage{amsfonts}
\usepackage{bbm}
\usepackage{babel}
\usepackage{xcolor}
\usepackage{verbatim}
\usepackage{mathrsfs}% \usepackage[utf8]{inputenc}
\usepackage{amsmath,amssymb,mathrsfs}%showkeys}
\usepackage{hyperref}
\everymath{\displaystyle}
\usepackage{amsmath,amssymb}
\usepackage{mathtools}
\usepackage{empheq}
\usepackage[T1]{fontenc}
\usepackage{lmodern}
\usepackage{amsmath,amssymb,mathtools}
\usepackage{booktabs}
\newcommand{\corr}[2]{\big\langle #1\,#2\big\rangle}
\newcommand{\tp}[1]{\tau^{-2\Delta-#1}}
\newcommand{\p}{\partial}
\newcommand{\td}{\tilde{\Delta}}
\newcommand{\Dt}{\mathcal{D}}

\newtheorem{lemma}{Lemma}
\newcommand{\be}[1]{ \begin{equation}\label{#1} }
\newcommand{\ee}{\end{equation}}
\newcommand{\bea}[1]{\begin{eqnarray}\label{#1} }
\newcommand{\eea}{\end{eqnarray}}
\newcommand{\oldeq}[1]{(#1)}          % reference to an equation elsewhere in the paper

\newcommand{\bes}{\begin{subequations}}
\newcommand{\ees}{\end{subequations}}

\title{\bf Galilean Kalb--Ramond Field}
\author[*]{Aditya Mehra} \author{\\}
\affiliation[*]{Department of Physics and Electronics, Christ University, Bengaluru, India, 560029\\}
\emailAdd{aditya.singh@christuniversity.in}

\abstract{
We construct Galilean limits of the free Kalb--Ramond two-form gauge field and analyze their symmetries using two different procedures. An In\"on\"u--Wigner contraction of the $D$-dimensional equations of motion results in two inequivalent electric and magnetic limits. In both cases the equations are invariant under the complete infinite-dimensional Galilean conformal algebra (GCA) only at $D=6$, the critical dimension of the relativistic theory. The second procedure is a null reduction of the $(D+1)$-dimensional theory which yields a local action for a four-field multiplet. In the zero-momentum sector, the resulting action is invariant under translations, rotations, Galilean boosts and an anisotropic dilatation with dynamical exponent $z$. However, in contrast to Galilean electrodynamics, it is not invariant under the special conformal generators of the GCA (and its higher modes). We attribute this to the vector residing at the bottom of the boost chain. If the parent theory is taken to be the six-dimensional conformal one, the resulting five-dimensional reduced action is invariant under the Schr\"odinger special conformal transformation with $z=2$ which is derived from the relativistic conformal group. In light-cone gauge, the non-zero null momentum sectors reduce to free Schr\"odinger fields. We obtain the gauge-fixed propagator, which is instantaneous, and compare the free-field correlators with the two-point functions allowed by the Ward identities of both.}

\begin{document}
\maketitle
\section{Introduction}
Symmetry principles constitute the sharpest tool available for organizing field theories, and the conformal symmetry \cite{DiFrancesco:1997nk} in particular has proved crucial in the most distant areas ranging from critical phenomena, string theory to the gauge/gravity correspondence. The real power of the conformal invariance, however, turns out to be dimension dependent: in two dimensions the algebra enhances to two copies of the infinite-dimensional Virasoro algebra and this infinite enhancement is what renders two-dimensional conformal field theories tractable without a need of a Lagrangian. In $D>2$ the relativistic conformal algebra $so(D,2)$ remains finite dimensional.

It is therefore quite remarkable that the situation improves, rather than the other way around, upon taking a non-relativistic limit \cite{Bergshoeff:2022eog,Hartong:2022lsy,Oling:2022fft}. The Inönü-Wigner contraction \cite{Inonu:1953sp} of the relativistic conformal algebra indeed gives a finite Galilean conformal algebra (fGCA), which has the same number of generators as its parent, but which can be naturally lifted to an infinite-dimensional algebra in any spacetime dimension \cite{Bagchi:2009my,Bagchi:2009pe,Duval:2009vt,Martelli:2009uc,Chen:2021xkw,Gupta:2020dtl,Bagchi:2016geg}. Whether this lift is just a curiosity, or whether it has a dynamical realization \cite{Bagchi:2022emh,Bidussi:2021nmp,Bagchi:2022eui2,Bagchi:2016bcd}, is a question that can only be answered on a case by case basis. The first example of higher-dimensional Galilean conformal symmetry was free Galilean electrodynamics: the electric and the magnetic limits of Maxwell theory both admit spacetime Galilean conformal symmetry, obtained by contracting the spacetime coordinates and simultaneously performing two different scalings of the gauge potential; the resulting equations of motion were invariant under the full infinite-dimensional GCA in $D=4$ \cite{Bagchi:2014ysa, LeBellac:1973unm,Festuccia:2016caf}. The same enhancement of symmetry occurs upon coupling to massless matter, and in the non-abelian case as well \cite{Bagchi:2015qcw, Bagchi:2017yvj, Mehra:2021sfx,Chapman:2020vtn,Baiguera:2023fus}. More recently, the technique of null reduction where we reduce a relativistic theory in $D+1$ dimensions along a null isometry \cite{Duval:1984cj,Chen:2023pqf,Lambert:2020zdc}. It has produced actions for these Galilean gauge theories \cite{Julia:1994bs, Festuccia:2016caf, Bagchi:2022eui,Saha:2025jsa}, at the price of extra fields, and a Hamiltonian analysis showed that the generators which do not act as symmetries of the action can however appear as Hamiltonian generators on phase space, with a state dependent central extension \cite{Banerjee:2019axy}.

All the examples are given by the one-form gauge potential, but, clearly, the next level of the hierarchy is also not exhausted: a free $p$-form gauge field in $D$ dimensions is scale-invariant for any $D$, but only conformally invariant in the critical dimension
\bea{}
D_c \;=\; 2(p+1) ,
\eea
which is the textbook statement for Maxwell theory, $p=1$, $D_c=4$ \cite{Jackiw:2011vz,ElShowk:2011gz, Nakayama:2013is}. The next member of the hierarchy is the Kalb--Ramond (KR) two-form $B_{\mu\nu}$, with $p=2$ and $D_c=6$. The KR field is not only the next member of the formal hierarchy, but also is an important ingredient of string theory: it is coupled to the string worldsheet by $\int_\Sigma B_{\mu\nu}\,dx^\mu\wedge dx^\nu$ \cite{Kalb:1974yc}, similarly to how the Maxwell potential is coupled to a charged particle, and is part of the massless NS--NS sector of string theory \cite{Oling:2022fft,Harmark:2017rpg}. The KR field, however, also provides the example with a genuine structure: its gauge symmetry $\delta B_{\mu\nu}=\partial_\mu\theta_\nu-\partial_\nu\theta_\mu$ is reducible. Furthermore, at $D=6$ the two-form is dual to another two-form, hence the critical dimension is the unique self-dual rank \cite{Aharony:1997an,Lambert:2020zdc,Lambert:2019jwi}; one would expect the electric--magnetic exchange to organise the Galilean theory similarly to how it does in the Maxwell case, but there is nothing analogous to GED \cite{Bagchi:2014ysa,Mehra:2021sfx}.

In this paper, we build the Galilean (non-relativistic) limit of Kalb--Ramond theory and look into the symmetries. We discuss two methods. The first is the contraction method. Here, we take the scaling of space-time indices and also of the field. It gives rise to two limits, namely, the electric and magnetic limits \cite{LeBellac:1973unm,Bagchi:2014ysa}. In both limits, we take the equations of motion and look at the Galilean conformal algebra as a symmetry. The equations turn out to be invariant under the infinite-dimensional GCA precisely in $D=6$, the critical dimension of the relativistic theory. 

We turn our attention to the second approach, called null reduction, which starts with a Lagrangian in $D+1$ dimensions and yields a $D$-dimensional Galilean Lagrangian. In the zero-momentum sector, the reduced action is invariant under Galilean translations, rotations, boosts, and dilatations, but not under the special conformal generators of the GCA or its higher modes. This is in contrast to Galilean electrodynamics, for which the null-reduced action retains the finite GCA in $D=4$ \cite{Bagchi:2022eui}. We show that the discrepancy is structural. For the one-form, the field at the base of the boost chain is the scalar $A_u$, and the exception was a total derivative. For the two-form, it is the vector $\psi_i=B_{ui}$, and the obstruction $\Psi_{lj}\psi_j$ does not share this property. We then wonder which non-relativistic conformal symmetry is \emph{inherited} derived from the parent conformal group. The generators preserving the reduction constitute the Schr\"odinger algebra, enlarged by a second dilatation unspecifying the dynamical exponent $z$. In the case of the parent critical theory in six dimensions, the five-dimensional reduced action is, of course, invariant under the Schr\"odinger special conformal transformation with $z=2$, and the sectors with non-zero null momentum reduce to free Schr\"odinger fields. We also compute the two-point functions. The contraction method makes it possible to use all symmetries, putting strong restrictions on the correlators. The null reduction method, by contrast, allows the boost identity to pass informations up the nilpotent ladder $E\to\{B,\chi\}\to\psi\to0$. We further obtain the gauge-fixed propagator of the null-reduced theory which is independent of the frequency, hence the free-field correlators have support at coincident times.

%The second approach is the null-reduction. In this method, we start from a Lagrangian in $D+1$ dimensions and end up with a $ D$-dimensional Lagrangian that is also Galilean in nature. Here an interesting twist takes place: the action is invariant under Galilean translations, boosts, rotations and dilatations. But it is not invariant under special conformal and higher modes. This is because of the presence of the $\Psi_{lj}\psi_j$ term in the solution. Finally, we calculate the two-point functions for both methods \cite{Bagchi:2009pe,Chen:2021xkw}. In the contraction method, we have all the symmetries. Because of that, the two-point functions get more constrained. In the null reduction method, the boost identity propagates data up a nilpotent ladder $E \to \{B,\chi\} \to \psi \to 0$, leaving a new constant at each level, whereas the special conformal identity, serves as a projector killing the boost--inert normalisation and all polynomial tails, leaving the ultra--local power law characteristic of a Galilean CFT. 

The flow of the paper is as follows. In Sec\eqref{sec:relativistic}, we start with a review of the relativistic conformal structure of the free KR theory. We locate the $(D-6)$ obstruction in the transformation of the field strength and in the trace of the stress tensor. In Sec\eqref{sec:nrcs}, we set up the GCA and look at its representation theory. We move on to see the two different methods of taking the Galilean limit on Kalb-Ramond theory. One is the contraction method (Sec\eqref{sec:contraction-invariance}), where we establish $D=6$ as the critical dimension in both limits, and the second one is the null reduction (Sec\eqref{sec:nullreduction}). 
In Sec.~\ref{sec:nullred} we construct the Galilean KR action by null reduction, analyse its invariance and trace the loss of the GCA special conformal generators to the structure of the boost chain. We then identify the symmetries inherited from the parent conformal group, which give Schr\"odinger invariance in $D=5$, and the free Schr\"odinger fields of the sectors with non-zero null momentum. Finally, in Sec\eqref{sec:corre-function}, we focus on the correlation functions in both frameworks. We close in Sec\eqref{sec:summary} with a summary and a list of future directions, including the canonical analysis and the possible role of this multiplet in the tensionless string.

\section{Conformal invariance and the Kalb--Ramond field}
\label{sec:relativistic}
The Kalb-Ramond (KR) field, $B_{\mu\nu} = -B_{\nu\mu}$, shows up as the rank-two entry in the family of $p$-form gauge potentials. It was first introduced because it naturally couples to a string’s worldsheet by the term $\int_\Sigma B_{\mu\nu} dx^\mu \wedge dx^\nu$, similar to as the Maxwell potential couples to a particle’s worldline \cite{Kalb:1974yc}. It’s a standard piece of the massless (NS–NS) sector in string theory. In $D$ dimensions, the KR field is actually the next rung after Maxwell theory in a classic sequence that explores the relation between scale invariance and conformal invariance \cite{Jackiw:2011vz,ElShowk:2011gz}. In general, a free $p$-form is scale invariant for any $D$, but conformal invariance only appears in a special “critical” dimension:
\be{}\label{eq:Dc}
D_c = 2(p+1),
\ee
So for the Maxwell field ($p = 1$), the critical dimension is $D_c = 4$, and for the KR field ($p = 2$), it is $D_c = 6$.
In this section, we will lay out the relativistic conformal structure of the free KR theory. We will see how both the potential and the field strength transform, explain where the “obstruction” terms (proportional to $D-6$) come from when we look at the invariance of KR theory under special conformal transformation (SCT), write down the trace of the stress tensor, and show the two-point functions right at criticality. 
These relativistic facts set the foundation for everything else to come: The scaling dimension $\tilde{\Delta} = \frac{D-2}{2}$, along with the boost and spin data for the Galilean multiplet, descend from this setup when you take the contraction. The critical dimension \eqref{eq:Dc} will show up, in Galilean setup, as the dimension where the limiting theories gets the infinite-dimensional symmetry. All coefficient identities stated here have been verified by
explicit component-wise symbolic computation in $D=4,5,6,7$.

%All coefficient identities quoted in this section have been verified by
%explicit componentwise symbolic computation in $D=4,5,6,7$.%
%\footnote{Script \texttt{KR\_rel\_verify.py}; the checks cover
%eqs.~\eqref{eq:dHfull}, \eqref{eq:obstruction}, \eqref{eq:trace}, the space--time splits
%\eqref{eq:EOMsplit}--\eqref{eq:bianchisplit}, the counting \eqref{eq:dof}, conservation of
%the two-point function \eqref{eq:HH} at $D=6$ and its failure at $D=5$, and, as a
%convention anchor, the Maxwell relations (2.13)--(2.14) of \cite{BBM}.}
\subsection{Conformal transformations of primary fields}
\label{sec:primaries}
In this paper, we adopt the metric signature $\eta_{\mu\nu}=\mathrm{diag}(-,+,\dots,+)$,
$\mu,\nu=0,\dots,D-1$. The conformal algebra in $D$ dimensions, is isomorphic to $\mathfrak{so}(D,2)$. It consists of translations $(\tilde P_\mu)$, Lorentz generators $(\tilde J_{\mu\nu})$, the dilatation $(\tilde D)$, and the special conformal transformations $(\tilde K_\mu)$. The generators realised as vector fields are given as
\bea{}\label{eq:relgen}&&
\tilde{P}_{\mu}=\partial_{\mu},~
\tilde{J}_{\mu\nu}=x_{\mu}\partial_{\nu}-x_{\nu}\partial_{\mu},~
\tilde{D}=-x^{\mu}\partial_{\mu},~
\tilde{K}_{\mu}=-\big(2x_{\mu}x^{\nu}\partial_{\nu}-x^{\nu}x_{\nu}\partial_{\mu}\big),
\eea
The commutators are given by
\bea{}\label{eq:relcomms}&&
[\tilde P_i,\tilde B_j]=-\delta_{ij}\tilde H,~
[\tilde B_i,\tilde B_j]=\tilde J_{ij},~
[\tilde K_i,\tilde B_j]=\delta_{ij}\tilde K,~
[\tilde K_i,\tilde P_j]=2\tilde J_{ij}+2\delta_{ij}\tilde D,
\eea{}
with $\tilde B_i\equiv\tilde J_{0i}$ denotes the Lorentz boosts and
$(\tilde H,\tilde K)\equiv(\tilde P_0,\tilde K_0)$.
The algebra \eqref{eq:relcomms} acts on a multi-component primary field
$\Phi(x)$ of scaling dimension $\td$ and spin matrix $\Sigma_{\mu\nu}$ as
\bes{}\label{eq:SCT}
\bea{}&&
\tilde\delta^{\rm trans}_\mu\Phi = \p_\mu\Phi, \qquad
\tilde\delta^{\rm Lorentz}_{\mu\nu}\Phi
 = \big(x_\mu\p_\nu-x_\nu\p_\mu+\Sigma_{\mu\nu}\big)\Phi,
\label{eq:poincare}\\&&
\tilde\delta^{\rm scale}\Phi = \big(x\!\cdot\!\p+\td\,\big)\Phi,
\label{eq:scale}\\&&
\tilde\delta^{\rm SCT}_\sigma\Phi
 = \big(2x_\sigma(x\!\cdot\!\p)-x^2\p_\sigma+2\td\,x_\sigma
   -2x^\tau\Sigma_{\tau\sigma}\big)\Phi,
\eea
\ees
where the canonical dimension is fixed by taking the scale invariance on the kinetic term of a two-derivative action. It comes out to be
\be{}
\td=\frac{D-2}{2}\,.
\label{eq:canonical}
\ee
Since, we will be dealing with the antisymmetric rank-two tensor representation, the spin matrix acts as
\bea{}\label{eq:spin2form}
\Sigma_{\tau\sigma}\,B_{\nu\lambda}
 =\eta_{\nu\tau}B_{\sigma\lambda}-\eta_{\nu\sigma}B_{\tau\lambda}
 +\eta_{\lambda\tau}B_{\nu\sigma}-\eta_{\lambda\sigma}B_{\nu\tau}\,,
\eea
\subsection{The free Kalb--Ramond theory}
\label{sec:freeKR}
The dynamical variable is the two-form potential $B_{\mu\nu}$ for the Kalb-Ramond theory, which has a totally antisymmetric field strength and a Lagrangian given by
\bea{}\label{eq:lagrangian}
H_{\mu\nu\lambda}=\p_\mu B_{\nu\lambda}+\p_\nu B_{\lambda\mu}+\p_\lambda B_{\mu\nu},
\qquad
\mathcal{L}=-\tfrac16\,H_{\mu\nu\lambda}H^{\mu\nu\lambda}.
\eea
and the theory is a gauge theory,
\bea{}\label{eq:gauge}
\delta B_{\mu\nu}=\p_\mu\theta_\nu-\p_\nu\theta_\mu\,,
\eea
with a feature missing from Maxwell theory: the gauge symmetry is reducible. The parameters $\theta_\mu$ and $\theta_\mu+\p_\mu\varphi$ create the same transformation \eqref{eq:gauge}, so there is a gauge symmetry of the gauge symmetry (``gauge-for-gauge''), with the scalar parameter $\varphi$. The field strength fulfills
\bea{}\label{eq:bianchi}&&
\text{Equations of motion:}\qquad \p^\mu H_{\mu\nu\lambda}=0,
\label{eq:EOM}\\&&
\text{Bianchi identity:}\label{eq:EOM1}\qquad 
\p_\sigma H_{\mu\nu\lambda}-\p_\mu H_{\nu\lambda\sigma}
+\p_\nu H_{\lambda\sigma\mu}-\p_\lambda H_{\sigma\mu\nu}=0,
\eea
the first comes from \eqref{eq:lagrangian}, and the second identically from $H=dB$.
The reducibility shows its impact in counting the degrees of freedom. The potential has $\binom{D}{2}$ components. Each of the $D$ gauge parameters seems to remove two. However, the parameters are redundant, and the correct (ghost-for-ghost) counting is the alternating sum
\bea{}\label{eq:dof}
N_{\rm dof}
=\binom{D}{2}-2\binom{D}{1}+3\binom{D}{0}
=\frac{(D-2)(D-3)}{2}=\binom{D-2}{2},
\eea
which represents the dimension of the antisymmetric tensor representation of the little group $SO(D-2)$. A light-cone analysis supports this. Notice the coefficient $+3$, not $+2$: a naive subtraction of the $D-1$ "effective" parameters undercounts by one. This discrepancy corresponds to the gauge-for-gauge mode.
Numerically, $N_{\rm dof}=1,3,6$ for $D=4,5,6$. The single degree of freedom in $D=4$ mirrors the well-known duality of the two-form to a massless scalar in that dimension. More generally, the equation of motion $d{\star}H=0$ implies ${\star}H=d\Lambda$ locally, where $\Lambda$ is a $(D-4)$-form. Thus, the free KR field is dual to a scalar in $D=4$, a vector in $D=5$, and to another two-form in $D=6$: the critical dimension \eqref{eq:Dc} is the unique self-dual rank. This suggests an electric-magnetic exchange structure that will run throughout the Galilean theory. It also clarifies why the key dimension for the KR story is $6$ and not $4$: in $D=4$, the two-form only captures the physics of a free scalar.
\subsection{Space--time decomposition}
\label{sec:split}
In the nonrelativistic limits, it is useful to record the theory in space--time split form, $\mu=(0,i)$ with $i=1,\dots,d$, $d=D-1$, $x^0=t$. The potential breaks down into an $SO(d)$ vector and an antisymmetric tensor,
\bea{}\label{eq:EBdef}
E_i\equiv B_{0i}\,~\text{and},~ B_{ij}\,,
\eea
which is the direct counterpart of $(A_0,A_i)$ in electrodynamics, and the KR field-strength
components are
\bea{}\label{eq:Hsplit}
H_{0ik}=\p_t B_{ik}+\p_k E_i-\p_i E_k\,,\qquad
H_{ijk}=\p_i B_{jk}+\p_j B_{ki}+\p_k B_{ij}\,.
\eea
Breaking the equations of motion \eqref{eq:EOM} into the components $(\mu=0,i)$ gives a constraint and an evolution equation,
\be{}
\p^i H_{0ik}=0 \quad (\nu\lambda=0k),
\qquad\qquad
\p^i H_{ijk}-\p_t H_{0jk}=0 \quad (\nu\lambda=jk),
\label{eq:EOMsplit}
\ee
These are the generalizations to the Gauss constraint and the Amp\`ere law. The constraint equation is
itself reducible to $\p^k\big(\p^i H_{0ik}\big)\equiv0$ identically by antisymmetry, in the canonical formalism, the reducibility of the gauge symmetry
\eqref{eq:gauge}. The Bianchi identity \eqref{eq:EOM1} breaks down into the analogues of
the Faraday law and of $\nabla\!\cdot\!B=0$,
\be{}\label{eq:bianchisplit}
\p_t H_{ijk}=\p_i H_{0jk}+\p_j H_{0ki}+\p_k H_{0ij}\,,
\qquad\qquad
\p_{[i}H_{jkl]}=0\,.
\ee
The equations of \eqref{eq:EOMsplit}, and identities \eqref{eq:bianchisplit}, are the
expressions whose leading behaviour under the two scalings of Sec.~3 defines the limits in the Galilean theories.
\subsection{Transformation of the field strength under Conformal Symmetry}
\label{sec:critical}
Under scale transformation, $B_{\mu\nu}$ field changes as a primary of dimension \eqref{eq:canonical}
and the field strength, as a covariant object of dimension
$\td+1=\tfrac{D}{2}$:
\bea{}\label{eq:scaleH}
\tilde\delta^{\rm scale}H_{\mu\nu\lambda}
=\Big(x\!\cdot\!\p+\tfrac{D}{2}\Big)H_{\mu\nu\lambda}\,=\Big(x\!\cdot\!\p+\td+1\Big)H_{\mu\nu\lambda}.
\eea
Since, the  Lagrangian $\mathcal{L}$ then carries weight $D$, the action $S=\int d^{D-1}x dt \mathcal{L}$ is scale invariant in any dimension just like free Maxwell theory. The Special conformal transformations (SCT) are a bit different. Applying  \eqref{eq:SCT} to $B_{\mu\nu}$ and taking the curl, we get after doing a lengthy calculations
\bea{}\label{eq:dHfull}
\tilde\delta^{\rm SCT}_\sigma H_{\mu\nu\lambda}
=\Xi_\sigma H_{\mu\nu\lambda}
+(D-6)\big(\eta_{\sigma\mu}B_{\nu\lambda}
          +\eta_{\sigma\nu}B_{\lambda\mu}
          +\eta_{\sigma\lambda}B_{\mu\nu}\big),
\eea
where
\bea{}\label{eq:XiH}
\Xi_\sigma H_{\mu\nu\lambda}
=\Big(2x_\sigma(x\!\cdot\!\p)-x^2\p_\sigma+D\,x_\sigma\Big)H_{\mu\nu\lambda}
-2x^\tau\,\Sigma^{(3)}_{\tau\sigma}H_{\mu\nu\lambda}
\eea
is the transformation $H$ under SCT as it would behave as a primary three-form of dimension
$\tfrac{D}{2}$, and with spin $\Sigma^{(3)}$ action analogous to
\eqref{eq:spin2form}. 
The field strength is not a conformal primary due to a term that depends on the bare potential, which turns off only in the critical dimension. When we apply $\p^\mu$ and use the equations of motion, the primary-like part adds terms proportional to $\p^\mu H_{\mu\nu\lambda}$ and its derivatives. What remains on shell is
\bea{}\label{eq:obstruction}
\p^\mu\,\tilde\delta^{\rm SCT}_\sigma H_{\mu\nu\lambda}\,\Big|_{\rm on\;shell}
=(D-6)\Big(\p_\nu B_{\sigma\lambda}-\p_\lambda B_{\sigma\nu}
-\eta_{\sigma\nu}\,\p^\mu B_{\mu\lambda}
+\eta_{\sigma\lambda}\,\p^\mu B_{\mu\nu}\Big)
\;\neq\;0,
\eea
for $D\neq6$. From above, we can put the result as
\begin{quote}
The free Kalb-Ramond theory is conformally invariant only when $D=6$. For $D\neq6$, it is scale invariant, but not conformally invariant.
\end{quote}
\subsection{Stress tensor and correlation functions}
\label{sec:stress}
The gauge-invariant symmetric stress tensor, derived by varying \eqref{eq:lagrangian} with respect to the metric, is given by 

\begin{equation}
T_{\mu\nu}=H_{\mu\alpha\beta}H_\nu{}^{\alpha\beta} - \tfrac16\,\eta_{\mu\nu}\,H_{\rho\sigma\lambda}H^{\rho\sigma\lambda}, 
\qquad 
T^\mu{}_\mu=-\frac{D-6}{6}\,H_{\mu\nu\lambda}H^{\mu\nu\lambda}, 
\label{eq:trace}
\end{equation}
and it remains conserved on shell. The trace vanishes identically in the critical dimension. In other cases, it is proportional to the Lagrangian itself. Since $H^2$ is not a total derivative, no improvement term can eliminate it. The reasoning from \cite{ElShowk:2011gz} applies directly to the two-form. Thus, the non-vanishing trace at \( D\neq6 \) represents the stress-tensor aspect of the obstruction \eqref{eq:obstruction}.

At the critical dimension, the conformal two-point functions take the usual primary form. For the potential, in a covariant gauge with $\td=2$ at $D=6$, we have
\bea{}\label{eq:BB2pt}
\big\langle B_{\mu\nu}(x)\,B_{\rho\sigma}(0)\big\rangle
=\frac{\eta_{\mu\rho}\eta_{\nu\sigma}-\eta_{\mu\sigma}\eta_{\nu\rho}}{(x^2)^{\td}}
\;+\;\text{gauge dependent terms},
\eea
Because it depends on the gauge, it is not physical. However, it generates the gauge-invariant correlator of field strengths. It is given by
\bea{}\label{eq:HH}
\big\langle H_{\mu\nu\lambda}(x)\,H_{\rho\sigma\tau}(0)\big\rangle
=\frac{1}{(x^2)^{D/2}}
\sum_{\pi\in S_3}\mathrm{sgn}(\pi)\,
I_{\mu\pi(\rho)}\,I_{\nu\pi(\sigma)}\,I_{\lambda\pi(\tau)}\,,~
I_{\mu\nu}=\eta_{\mu\nu}-\frac{2x_\mu x_\nu}{x^2},
\eea
This expression shows the three-index antisymmetrized product of inversion tensors. To stay consistent with the equations of motion, we need $\p^\mu\langle H_{\mu\nu\lambda}H_{\rho\sigma\tau}\rangle=0$ at separated points. This condition is satisfied by \eqref{eq:HH} exactly at $D=6$ and does not hold away from it. We will not explore the relativistic correlators further here. Their Galilean counterparts, built from the contracted symmetry algebra, will be discussed further in the paper.

\section{Non-relativistic conformal symmetry and the Kalb--Ramond multiplet}
\label{sec:nrcs}

Several different ideas about conformal symmetry exist in the non-relativistic setting. The one relevant for the massless (gapless) two-form is the Galilean conformal algebra (GCA). This algebra is derived from an Inönü-Wigner contraction of the relativistic conformal algebra, similar to how the Galilean algebra comes from the Poincaré algebra. Throughout this paper, the Galilean theory takes place in $D$ spacetime dimensions: a time coordinate $t$ and $(D-1)$ spatial coordinates $x^{i}$, where $i=1,\dots,D-1$. The rotation group is $SO(D-1)$.

We will discuss two constructions. The first involves the contraction of the relativistic Kalb-Ramond (KR) theory in $D$ dimensions. This contraction affects the equations of motion and results in a two-field Galilean multiplet. The second construction is the null reduction of the relativistic theory in $D+1$ dimensions. This produces an action and a larger, four-field multiplet. The two constructions single out the critical dimension in different ways. The contracted equations are invariant under the infinite-dimensional GCA precisely at $D=6$. In the null reduction, the symmetries inherited from the parent conformal group appear when the parent theory is critical, $D+1=6$, and they form a Schr\"odinger rather than a Galilean conformal algebra. The representation theory in Sec(\ref{sec:rep}) must address both multiplets at the same time. Therefore, we first establish the null-reduction ansatz and then determine the action of the algebra afterwards, once all the fields are presented.

\subsection{The Galilean conformal algebra}
\label{sec:contraction}
The non-relativistic limit is established at the level of spacetime, and also take \(c=1\). This is done through the anisotropic rescaling given by
\bea{}\label{eq:contraction}
x^{i} \to \epsilon x^{i},\qquad
t \to t,\qquad \epsilon \to 0,
\eea
This rescaling is the same as letting \(c\) approach infinity, which leads to \(\partial_{i} \to \epsilon^{-1} \partial_{i}\). By contracting \eqref{eq:relgen} in this manner, we find
\bes{}
\bea{}\label{eq:gcagen}
H = -\partial_{t},  \bar{D} = -\big(t\partial_{t}+x^{i}\partial_{i}\big),  K = -\big(t^{2}\partial_{t}+2t x^{i}\partial_{i}\big), \\
P_{i} = \partial_{i},  B_{i} = t \partial_{i},  K_{i} = t^{2} \partial_{i}, \\
J_{ij} = x_{i} \partial_{j}-x_{j} \partial_{i},
\eea
\ees
where $( (H,\bar{D},K) )$ represent the Hamiltonian, dilatation, and temporal SCT, while \( (P_{i},B_{i},K_{i}) \) represent spatial translations, Galilean boosts, and spatial SCT. The non-zero commutators are
\bea{}\label{eq:fgca}
&&[J_{ij},J_{kl}]=\delta_{ik}J_{jl}-\delta_{jk}J_{il}-\delta_{il}J_{jk} +\delta_{jl}J_{ik},[J_{ij},X_{k}]=\delta_{ik}X_{j}-\delta_{jk}X_{i}, \nonumber\\
&&[H,B_{i}]=-P_{i},\qquad [B_{i},B_{j}]=0,\qquad [P_{i},B_{j}]=0, 
[\bar{D},H]=H,\quad [\bar{D},P_{i}]=P_{i}, \nonumber\\
&&[\bar{D},K]=-K,\qquad [\bar{D},K_{i}]=-K_{i},\qquad [K,H]=2\bar{D}, 
[K,P_{i}]=2B_{i},\quad [B_{i},K]=-K_{i}, \nonumber\\
&&[K_{i},H]=2B_{i},\qquad [B_{i},K_{j}]=0,\qquad [K_{i},P_{j}]=0, 
\eea
where \(X_{i}=(P_{i},B_{i},K_{i})\). These commutators in this collection are known as the finite Galilean conformal algebra (fGCA). Notably, the boosts become abelian; this property will be essential in Sec.~\ref{sec:rep}.

The generators arrange themselves as,
\bea{}\label{eq:LMdef}
L^{(n)}=-t^{n+1}\partial_{t}-(n+1)t^{n}x^{i}\partial_{i},\qquad
M^{(n)}_{i}=t^{n+1}\partial_{i},\qquad
J_{ij}=-(x_{i}\partial_{j}-x_j \partial_{i}),
\eea
with \(L^{(-1,0,1)}=(H,\bar{D},K)\) and \(M^{(-1,0,1)}_{i}=(P_{i},B_{i},K_{i})\). In this way, \eqref{eq:fgca} becomes
\bea{}\label{eq:igca}&&
[L^{(n)},L^{(m)}]=(n-m)L^{(n+m)}, 
[L^{(n)},M^{(m)}_{i}]=(n-m)M^{(n+m)}_{i}, 
[M^{(n)}_{i},M^{(m)}_{j}]=0, \nonumber\\&&
[L^{(n)},J_{ij}]=0, 
[J_{ij},M^{(n)}_{k}]=\delta_{ik}M^{(n)}_{j}-\delta_{jk}M^{(n)}_{i}.
\eea
Although derived for \(n=0,\pm1\), the brackets \eqref{eq:igca} hold for all \(n \in \mathbb{Z}\): the fGCA expands into an infinite-dimensional structure, a Witt algebra formed by the \(L^{(n)}\) along with an abelian ideal made up of the \(M^{(n)}_{i}\). We simply refer to this infinite extension as the GCA. The Witt subalgebra allows for the usual central extension, which we will see in the phase-space realization of the theory.

As discussed earlier, during the contraction, the relativistic two-form separates into an \(SO(D-1)\) vector and an antisymmetric tensor:
\be{}\label{eq:contmultiplet}
E_{i} \equiv B_{0i},\qquad B_{ij},
\ee
The limit remains consistent in two distinct ways, depending on which components are kept fixed as the light cone expands:
\bea{}\label{eq:scalings}
\text{Electric:}\quad B_{0i} \to B_{0i},\; B_{ij} \to \epsilon B_{ij};
\qquad
\text{Magnetic:}\quad B_{0i} \to \epsilon B_{0i},\; B_{ij} \to B_{ij},
\eea

\subsection{Geometric realisation}
The algebra \eqref{eq:igca} also has a geometric description. A flat Newton-Cartan structure \((\mathcal{G}, h, \tau, \Gamma)\) includes a degenerate spatial metric \(h^{\mu\nu}\), a clock one-form \(\tau = \tau_{a}dx^{a}\) with \(\tau_{a}h^{ab} = 0\), and a connection \(\Gamma\) that has no torsion. Its conformal isometries come from vector fields that satisfy the equations
\bea{}
\mathcal{L}_{\xi}h^{\mu\nu} = \lambda h^{\mu\nu},\qquad \mathcal{L}_{\xi}\tau_{a} = \mu \tau_{a},\qquad \lambda + N\mu = 0,
\eea
These generate the conformal Galilei algebra of level \(N\) with a dynamical exponent \(z = 2/N\). The focus here is on the case \(N = 2\), which means \(z = 1\). For this case, \(\mathfrak{cgal}_{2}(\mathcal{G}, h, \tau) = \mathfrak{gca}(d)\) corresponds exactly to the infinite-dimensional GCA of \eqref{eq:igca}. If we also require that the transformations keep the projective structure of the flat connection, it reduces to the finite GCA \eqref{eq:fgca}. Unless stated otherwise, symmetry statements refer to $z=1$; the Schr\"odinger case $z=2$ appears in Secs.~5.6--5.7.
\subsection{Null reduction}
\label{sec:nullred}
The contraction \eqref{eq:contraction} operates on the equations of motion but does not produce an action principle. The alternative method that does produce one is null reduction. This approach generates a $D$-dimensional Galilean invariant theory from a $(D+1)$-dimensional relativistic theory by reducing along a null direction. In this section, we will outline the process, including the field content it creates, which the representation theory must address. We will implement this for the Kalb–Ramond field.

We start with $(D+1)$-dimensional Minkowski space, represented by the equation $ds^{2}=-(dx^{0})^{2}+(dx^{1})^{2}+\delta_{ij}dx^{i}dx^{j}$. We then switch to light-cone coordinates defined by
\begin{equation}
u=\tfrac{1}{\sqrt2}\left(x^1-x^0\right),\qquad t=\tfrac{1}{\sqrt2}\left(x^0+x^1\right),
\end{equation}
so that $-(dx^0)^2+(dx^1)^2=2\,du\,dt$
which transforms the metric into
\bea{}\label{eq:lcmetric}
ds^{2}=\eta_{\tilde\mu\tilde\nu}dx^{\tilde\mu}dx^{\tilde\nu}
=2\,du\,dt+\delta_{ij}dx^{i}dx^{j},
\qquad
\eta_{ut}=\eta_{tu}=1,\quad \eta_{ij}=\delta_{ij},
\eea
where $\tilde\mu=(u,t,i)$ and $i=1,\dots,D-1$. The inverse metric gives $\eta^{ut}=\eta^{tu}=1$ and $\eta^{ij}=\delta^{ij}$. This means raising an index swaps $u$ and $t$ while leaving $i$ unchanged, which we will use frequently. We then reduce along $u$: all fields no longer depend on $u$, so $\partial_{u}=0$. The coordinate $t$ now acts as Galilean time, and the $x^{i}$ become the spatial coordinates in the $D$-dimensional Galilean theory.

As a preliminary example, consider a free complex scalar in $D+1$ dimensions,
\bea{}
S=-\int d^{D+1}x\,\eta^{\tilde\mu\tilde\nu}\,\partial_{\tilde\mu}\bar\phi\,\partial_{\tilde\nu}\phi.\eea
Since $\partial_u$ is a null Killing vector, the momentum along $u$ is conserved and the field can be decomposed as $\phi=e^{imu}\varphi_m(t,x^i)$. For $m\neq0$ the cross terms $\partial_u\bar\phi\,\partial_t\phi+\partial_t\bar\phi\,\partial_u\phi$ give
\bea{}\label{eq:bargmann-scalar}
S_m=\int dt\,d^{D-1}x\,\Big[\,i m\big(\bar\varphi_m\partial_t\varphi_m-\varphi_m\partial_t\bar\varphi_m\big)-\partial_i\bar\varphi_m\,\partial_i\varphi_m\Big],
~ i\partial_t\varphi_m=-\frac{1}{2m}\nabla^2\varphi_m ,
\eea
the free Schr\"odinger field of mass $m$. This idea comes from the Bargmann (Eisenhart-Duval) construction, which has been discussed in various studies \cite{Duval:1984cj,Julia:1994bs}. The key point here is that the symmetries related to the parent theory that act on the null momentum \(P_u\) create what we call the Schrödinger algebra. This algebra possesses a dynamical exponent of $z=2$ \cite{Son:2008ye, Balasubramanian:2008dm, Duval:2008jg}. 

However, our focus is different \cite{Festuccia:2016caf,Mehra:2021sfx,Bagchi:2022eui}. We consider the zero-momentum sector, which reduces to the case $m=0$. In this context, we have $\partial_u \Phi = 0$. When dealing with scalar fields, things can get a bit tricky. The terms involving $\partial_t$ also include $\partial_u$, leading to a reduced Lagrangian that doesn’t feature any time derivatives, making time just a label in this setup.

On the other hand, gauge fields act differently because the null component of their potential maintains a time derivative through the field strength. For example, in Maxwell's theory, we find that $F_{ut} = -\partial_t A_u$, and reducing the expression results in the Lagrangian for Galilean electrodynamics. This is expressed as $\mathcal L_{\text{GED}} = \frac12 (\partial_t \phi)^2 + E_i \partial_i \phi - \frac14 W_{ij}W_{ij}$, where $\phi$ corresponds to $A_u$ and $E_i =(\p_t a_i - \p_i a_t)$ and $W_{ij}=(\p_i a_j - \p_j a_i)$. Here, $(a_t,a_i,\phi)$ are the Galilean gauge and scalar fields. \cite{Festuccia:2016caf,Mehra:2021sfx,Bagchi:2022eui}. We use this framework to ensure our conventions are consistent. We will later return to the sectors with $m \neq 0$ and examine their Schrödinger symmetry as well.

% This is the Bargmann (Eisenhart--Duval) construction \cite{}[25,28]. The symmetries of the parent theory that commute with the null momentum $P_u$ form the Schr\"odinger algebra, with dynamical exponent $z=2$ [S-Son, S-BM, S-DHH]. In this paper we instead work, following [19,22,29], in the zero-momentum sector $m=0$, i.e.\ $\partial_u\Phi=0$. For a scalar, this sector is degenerate: every term containing $\partial_t$ also contains $\partial_u$, so the reduced Lagrangian $-\partial_i\bar\varphi_0\,\partial_i\varphi_0$ has no time derivative and $t$ is only a label. Gauge fields behave differently, because the null component of the potential keeps a time derivative through the field strength. For Maxwell theory $F_{ut}=-\partial_t A_u$, and the reduction of $-\frac14 F^2$ gives the Galilean electrodynamics Lagrangian $\mathcal L_{\rm GED}=\frac12(\partial_t\phi)^2+E_i\partial_i\phi-\frac14 W_{ij}W_{ij}$ with $\phi\equiv A_u$ [19,22,29], which we use to check our conventions. For the two-form, the corresponding role is played by $H_{uti}=\partial_i\chi-\partial_t\psi_i$, see (5.4a). We return to the $m\neq0$ sectors of the two-form, and to their Schr\"odinger symmetry later in the paper.
For our specific case, the ansatz applies to the two-form $B_{\tilde\mu\tilde\nu}$, which splits into four $SO(D-1)$ multiplets:
\bea{}\label{eq:nullmultiplet}
\chi\equiv B_{ut}\ (\text{scalar}),\qquad
\psi_{i}\equiv B_{ui}\ (\text{vector}),\qquad
E_{i}\equiv B_{ti}\ (\text{vector}),\qquad
B_{ij}\ (\text{tensor}).
\eea
Comparing this with \eqref{eq:contmultiplet}, null reduction keeps the two fields from the contraction module and adds two more, $\chi$ and $\psi_{i}$, which arise from the null direction. These additional fields are essential for creating an action. As we will demonstrate, they also hinder the special conformal generators at the level of that action. By setting $\chi=\psi_{i}=0$, we narrow the four-field system down to the magnetic sector of the contraction.

\subsection{The Kalb-Ramond multiplet and its scale-spin representation}
\label{sec:rep}
We now write down the representation of \eqref{eq:igca} based on the two multiplets we just discussed:
\begin{enumerate}
    \item The two-field contraction module $\{E_{i},B_{ij}\}$ from \eqref{eq:contmultiplet},
    \item and the four-field null-reduction module $\{\chi,\psi_{i},E_{i},B_{ij}\}$ from \eqref{eq:nullmultiplet}.
\end{enumerate}
This serves as the technical foundation for every symmetry statement in the rest of the paper. We denote a generic member of the multiplet as $\Phi^{A}$.
 
We will now speak on the scaling dimension. Both multiplets share a single common scaling dimension  
\bea{}\label{eq:Delta}
\Delta=\frac{D-2}{2}.
\eea  
The contraction module inherits this directly. The Galilean dilatation $L^{(0)}$ is the undeformed contraction of $\tilde{D}$, denoted by $\bar{D}$. In contrast, for the null-reduction module, this is a result. The dilatation invariance of the reduced action determines the four dimensions independently, and at $z=1$, they reduce to the common value \eqref{eq:Delta}.

Moving on to understand the concept of primaries for our case. From \eqref{eq:igca}, we have $[L^{(0)},L^{(n)}]=-nL^{(n)}$ and $[L^{(0)},M^{(n)}_{i}]=-nM^{(n)}_{i}$. Thus, modes where $n>0$ lower the $L^{(0)}$ weight. A primary is defined by  
\bea{}
L^{(n)}\Phi=M^{(n)}_{i}\Phi=0 \qquad (n>0).
\eea 
Since $[L^{(0)},M^{(0)}_{i}]=[L^{(0)},J_{ij}]=0$ while $[M^{(0)}_{i},J_{ij}]\neq0$, primaries are characterized by the dilatation weight and an $SO(D-1)$ spin, where $[J_{ij},\Phi]=\Sigma_{ij}\triangleright\Phi$. Using the state-operator map, the data at the origin is given by  
\bea{}
[L^{(0)},\Phi]=\Delta\,\Phi,~
[L^{(1)},\Phi]=[M^{(1)}_{i},\Phi]=0,~
[L^{(-1)},\Phi]=\partial_{t}\Phi,~
[M^{(-1)}_{i},\Phi]=-\partial_{i}\Phi,
\eea 
and the spin operates in the standard tensorial manner,  
\bea{}\label{eq:spinaction}
[J_{ij},E_{m}]=\delta_{im}E_{j}-\delta_{jm}E_{i},\qquad 
[J_{ij},B_{mn}]=\delta_{im}B_{jn}-\delta_{jm}B_{in}
+\delta_{in}B_{mj}-\delta_{jn}B_{mi}.
\eea
\paragraph{Boosts:} All the model-dependent information is contained in the action of the boost $B_{k}=M^{(0)}_{k}$. Because boosts do not commute with rotations, we cannot read this action directly from \eqref{eq:spinaction}. Instead, it is restricted by the Jacobi identity for $\{J_{ab},B_{k},\Phi\}$. This identity states that $[J_{ab},B_{k}]=\delta_{ak}B_{b}-\delta_{bk}B_{a}$, which leads to the covariance condition 

\bea{}\label{eq:covcond}
[J_{ab},[B_{k},\Phi]]
=[B_{k},\Sigma_{ab}\triangleright\Phi]
+\delta_{ak}[B_{b},\Phi]-\delta_{bk}[B_{a},\Phi].
\eea

To solve \eqref{eq:covcond} for the two-form multiplet, we find that the only dimension-preserving, $SO(D-1)$-covariant options are captured by the single rule 
\bea{}\label{eq:boostrule}
[B_{k},\Phi^{A}]
=\delta^{E}_{A}\Big(a\,B_{mk}+r\,\delta_{km}\chi\Big)
+\delta^{B}_{A}\Big(b\,(\delta_{kn}E_{m}-\delta_{km}E_{n}) 
+s\,(\delta_{kn}\psi_{m}-\delta_{km}\psi_{n})\Big)
+c\,\delta^{\chi}_{A}\,\psi_{k},\nonumber\\
\eea
where $\Phi^{A}\in\{\chi,\psi_{i},E_{i},B_{ij}\}$. The selector $\delta^{X}_{A} \equiv 1$ if $\Phi^{A}=X$ and is zero otherwise. The indices $m,n$ are the free indices attached to $\Phi^{A}$. The constants $a,b$ control the $E\leftrightarrow B$ mixing that is common to both modules, while $r,s,c$ represent the additional constants present in the fields $\chi,\psi_{i}$ from \eqref{eq:nullmultiplet}. Each term separately satisfies \eqref{eq:covcond}. The dynamics then determine the values of the constants.

\paragraph{Contraction module:}In the contraction module, we only have $a,b$. The commutators that survive are
\bea{}\label{eq:abrule}
[B_{k},E_{m}]=a\,B_{mk},\qquad
[B_{k},B_{mn}]=b\,(\delta_{kn}E_{m}-\delta_{km}E_{n}).
\eea

These structures come from the Lorentz boost $\tilde{J}_{0k}$. Under this boost, a two-form transforms according to the spin part:  
$\Sigma_{0k}B_{\rho\sigma}=(\eta_{0\rho}B_{k\sigma}-\eta_{k\rho}B_{0\sigma})+(\eta_{0\sigma}B_{\rho k}-\eta_{k\sigma}B_{\rho0})$. If we let $E_{j}\equiv B_{0j}$ and $B_{j0}=-E_{j}$ (note that $\eta_{00}=-1$), we have  
\bea{} \label{eq:lorentzboost}  
\Sigma_{0k}E_{j}=\eta_{00}B_{kj}=B_{jk},\qquad  
\Sigma_{0k}B_{jl}=-\eta_{kj}B_{0l}-\eta_{kl}B_{j0}  
=\delta_{kl}E_{j}-\delta_{kj}E_{l},  
\eea
which corresponds to \eqref{eq:abrule} at $a=b=1$. However, a true \emph{Galilean} module must have commuting boosts, $[B_{i},B_{j}]=0$. Since $[[B_{i},B_{j}],E_{m}]=ab\,(\delta_{jm}E_{i}-\delta_{im}E_{j})$, this leads to  
\bea{}\label{eq:abzero} 
\boxed{\;ab=0\;}   
\eea
so that only one of the two mixings remains. The surviving one is determined by the field scalings \eqref{eq:scalings} that go with the contraction:  
\bea{}  
\text{Electric:}\;(a,b)=(0,1),  
\qquad  
\text{Magnetic:}\;(a,b)=(1,0).  
\label{eq:abfix}  
\eea 
Both cases satisfy \eqref{eq:abzero}, and the two limits can be swapped by $(a,b)\to(b,a)$. This swap reflects the representation-theoretic aspect of the electric-magnetic duality that appears throughout the paper, including in the invariance obstructions and the correlation functions. It is ultimately based on the self-duality of the two-form at the critical dimension.

\paragraph{Null-reduction module.}
For the four-field multiplet \eqref{eq:nullmultiplet} the boost is the light-cone one, $\delta x^{i}=v^{i}t$, $\delta u=-v_{i}x^{i}$, $\delta t=0$, with
$\omega_{it}=-\omega_{ti}=\delta_{ik}$. The spin variation
$\delta_{\mathrm{spin}}B_{\tilde\mu\tilde\nu}
=-\omega^{\tilde\rho}{}_{\tilde\mu}B_{\tilde\rho\tilde\nu}
-\omega^{\tilde\rho}{}_{\tilde\nu}B_{\tilde\mu\tilde\rho}$ then fixes
\bea{}\label{eq:nullconsts}
(a,b,r,s,c)=(1,0,-1,-1,-1),
\eea
that is, writing $S_{k}[\Phi]\equiv[B_{k},\Phi(0,0)]$,
\bea{}\label{eq:nullboost}
S_{k}[\chi]=-\psi_{k},~
S_{k}[\psi_{i}]=0,~
S_{k}[E_{i}]=B_{ik}-\delta_{ik}\chi,~
S_{k}[B_{ij}]=\delta_{ik}\psi_{j}-\delta_{jk}\psi_{i}.
\eea
It should be noted that $\psi_{i}$ is at the bottom of the
boost chain, $S_{k}[\psi_{i}]=0$: it will therefore transform purely orbitally.
Throughout the entire algebra, there is a fact that leads to a number of cancellations later on.

\paragraph{Transport off the origin.}
The action of the algebra at a generic point can be found by translating,
$\Phi(t,x)=U\Phi(0,0)U^{-1}$ with
$U=e^{A}$, $A=tL^{(-1)}-x^{j}M^{(-1)}_{j}$.
So that, we get
$[\mathcal{O},\Phi(t,x)]=U[U^{-1}\mathcal{O}U,\Phi(0,0)]U^{-1}$ 
with
$U^{-1}\mathcal{O}U=\sum_{p}\frac{(-1)^{p}}{p!}\mathrm{ad}^{p}_{A}(\mathcal{O})$
Using it, we have
\bea{}\label{symm}
\mathrm{ad}_{A}\big(M^{(n)}_{l}\big)=-(n+1)t\,M^{(n-1)}_{l},~
\mathrm{ad}_{A}\big(L^{(n)}\big)=-(n+1)t\,L^{(n-1)}+(n+1)x^{j}M^{(n-1)}_{j}.
\eea
Each $\mathrm{ad}_{A}$ lowers the mode number by one, so the series terminates;
acting on a primary only the modes $n-p=0$ (contributing $\Delta$ and $S_{k}$)
and the cases where $n-p=-1$ (the translations) remain. When we collect the terms,
\bes{}\label{eq:Mact}
\bea{}
\big[L^{(n)},\Phi(t,x)\big]
=\Big(t^{n+1}\partial_{t}+(n+1)t^{n}\big(x^{l}\partial_{l}+\Delta\big)\Big)\Phi
-n(n+1)t^{n-1}x^{k}\,S_{k}[\Phi],
\label{eq:Lact}\\[2pt]\label{eq:Mact1}
\big[M^{(n)}_{l},\Phi(t,x)\big]
=-t^{n+1}\partial_{l}\Phi+(n+1)t^{n}\,S_{l}[\Phi],
\eea\ees
where $S_{k}[\Phi]=U[M^{(0)}_{k},\Phi(0,0)]U^{-1}$ is given by
\eqref{eq:boostrule} with the relevant constants.
\paragraph{Action on the multiplet:}
Putting the boost data \eqref{eq:nullboost} into
\eqref{eq:Lact}--\eqref{eq:Mact1} and using
$\Dt^{(n)}\equiv t^{n+1}\partial_{t}+(n+1)t^{n}(x^{l}\partial_{l}+\Delta)$, the
spatial tower acts as
\bes{}\label{eq:Mchi}
\bea{}&&
\big[M^{(n)}_{l},\chi\big] = -t^{n+1}\partial_{l}\chi-(n+1)t^{n}\psi_{l},~
\big[M^{(n)}_{l},\psi_{i}\big] = -t^{n+1}\partial_{l}\psi_{i},\\&&
\big[M^{(n)}_{l},E_{i}\big] = -t^{n+1}\partial_{l}E_{i}
+(n+1)t^{n}\big(B_{il}-\delta_{il}\chi\big),\\&&
\big[M^{(n)}_{l},B_{ij}\big] = -t^{n+1}\partial_{l}B_{ij}
+(n+1)t^{n}\big(\delta_{il}\psi_{j}-\delta_{jl}\psi_{i}\big),
\eea\ees
and the temporal part as
\bes{}\label{eq:LB}
\bea{}&&
\big[L^{(n)},\chi\big] = \Dt^{(n)}\chi+n(n+1)t^{n-1}x^{k}\psi_{k},~
\big[L^{(n)},\psi_{i}\big] = \Dt^{(n)}\psi_{i},\\&&
\big[L^{(n)},E_{i}\big] = \Dt^{(n)}E_{i}
-n(n+1)t^{n-1}\big(x^{k}B_{ik}-x_{i}\chi\big),\\&&
\big[L^{(n)},B_{ij}\big] = \Dt^{(n)}B_{ij}
-n(n+1)t^{n-1}\big(x_{i}\psi_{j}-x_{j}\psi_{i}\big).
\eea\ees
The auxiliary vector $\psi_{i}$, being the bottom of the boost chain, transforms
purely orbitally, while $\chi$, $E_{i}$ and $B_{ij}$ carry algebraic spin pieces. Additionally, in the case of the contraction module one merely sets $\chi$ and $\psi_{i}$ equal to zero and uses
\eqref{eq:abrule} with the values \eqref{eq:abfix}, so that $S_{k}[E_{m}]=aB_{mk}$
and $S_{k}[B_{mn}]=b(\delta_{kn}E_{m}-\delta_{km}E_{n})$.
\section{The Galilean limit by contraction and its invariance}
\label{sec:contraction-invariance}
The representation theory has knowledge of the boost constants $(a,b)$ but does not care which particular values of these the dynamics picks. We provide the latter now directly, through a contraction of the free equations of motion. The free Kalb-Ramond field has \emph{two} nonequivalent Galilean limits, electric and magnetic, depending on which components of the two-form we keep in the limit of opening the light cone. We work out the equations of motion in these limits, determine the values of $(a,b)$ in each of these sectors, and prove that in both cases the infinite GCA is invariant precisely in $D=6$.
\subsection{Contraction and the two limits}
\label{sec:twolimits}
We use the spacetime contraction \eqref{eq:contraction},
$x^{i}\to\epsilon x^{i}$, $t\to t$, $\epsilon\to0$
(so $\partial_{i}\to\epsilon^{-1}\partial_{i}$), and combine it with the two scalings of the potential given in
\eqref{eq:scalings},
\bea{}\label{eq:scalings-recall}
\text{Electric:}\quad B_{0i}\to B_{0i},\; B_{ij}\to\epsilon B_{ij};
\qquad
\text{Magnetic:}\quad B_{0i}\to\epsilon B_{0i},\; B_{ij}\to B_{ij}.
\eea
In each case, retaining the leading term of \eqref{eq:EOMsplit} gives a consistent set of Galilean equations.
\paragraph{Electric limit:}
In this limit, $E_{i}=B_{0i}$ dominates. The equations of motion are
\bes{}\label{eq:elecEOM2}
\bea{}
\partial^{i}\partial_{i}E_{k}-\partial_{k}\,\partial^{i}E_{i}=0 ,
\label{eq:elecEOM1}\\
\partial^{i}\partial_{i}B_{jk}+\partial^{i}\partial_{j}B_{ki}
+\partial^{i}\partial_{k}B_{ij}
+\partial_{t}\big(\partial_{j}E_{k}-\partial_{k}E_{j}\big)=0 .
\eea\ees
The first is the contraction of the constraint $\partial^{i}H_{0ik}=0$.
The second is the contraction of the evolution equation, in which the
$B$-block and the curl of $E$ persist alongside each other.

\paragraph{Magnetic limit:}
In magnetic limit, the spatial block $B_{ij}$ dominates, and the roles are reversed,
\bes{}\label{eq:magEOM02}
\bea{}\label{eq:magEOM2}
&\partial^{i}\partial_{i}B_{jk}+\partial^{i}\partial_{j}B_{ki}
+\partial^{i}\partial_{k}B_{ij}=0 ,
\label{eq:magEOM1}\\
&\partial^{i}\partial_{i}E_{k}-\partial_{k}\,\partial^{i}E_{i}
-\partial_{t}\,\partial^{i}B_{ik}=0 .
\eea\ees
When we compare \eqref{eq:elecEOM2} with \eqref{eq:magEOM02}, the time
derivative attaches to the curl of $E$ in the electric limit and to the
divergence of $B$ in the magnetic sector. This is the first hint of the
electric--magnetic duality between the two limits.
\paragraph{Fixing the boost constants:}
If the contraction is applied to the conformal transformation of the fields (as opposed to being applied to the equations), then the representation constants (a,b) of \eqref{eq:abrule} are fixed. We get

\bea{}\label{eq:abfix-recall}
\text{Electric:}\;(a,b)=(0,1),
\qquad
\text{Magnetic:}\;(a,b)=(1,0),
\eea

Both satisfy the Galilean condition $ab=0$ of \eqref{eq:abzero}. In the
electric limit $E_{i}$ is boost-inert ($S_{k}[E_{m}]=0$) while $B_{ij}$
mixes into $E$; in the magnetic sector the pattern is exchanged,
$S_{k}[E_{i}]$ equals $B_{ik}$ and $B_{ij}$ is inert. We will use them to test the invariance of the equations under GCA.
\subsection{GCA invariance: Electric limit}
\label{sec:elecinv}

We write the electric equations \eqref{eq:elecEOM2} as
$T^{(1)}_{k}=0$ and $T^{(2)}_{jk}=0$, where
\bea{}\label{eq:elecT}
T^{(1)}_{k}=\partial^{i}\partial_{i}E_{k}-\partial_{k}\partial^{i}E_{i},
\qquad
T^{(2)}_{jk}=\partial^{i}H_{ijk}+\partial_{t}C_{jk},
\qquad
C_{jk}\equiv\partial_{j}E_{k}-\partial_{k}E_{j}.
\eea
To check for the invariance under infinite-GCA generators, we use the field action \eqref{eq:Mact} with the electric boost data
$(a,b)=(0,1)$. For simplification, we write
$\Dt^{(n)}\equiv t^{n+1}\partial_{t}+(n+1)t^{n}(x^{l}\partial_{l}+\Delta)$
and using the operator relations
$[\partial_{p},\Dt^{(n)}]=(n+1)t^{n}\partial_{p}$ and
$[\partial_{t},\Dt^{(n)}]=(n+1)t^{n}\partial_{t}
+n(n+1)t^{n-1}(x^{l}\partial_{l}+\Delta)$. Using it, under $M^{(n)}_l$, we get 
\bea{}\label{eq:elecM}
[M^{(n)}_{l},T^{(1)}_{k}]=-t^{n+1}\partial_{l}T^{(1)}_{k},
~
[M^{(n)}_{l},T^{(2)}_{jk}]=-t^{n+1}\partial_{l}T^{(2)}_{jk}
+(n+1)t^{n}\big(\delta_{lk}T^{(1)}_{j}-\delta_{lj}T^{(1)}_{k}\big).
\eea
These equations tell that they are invariant under $M^{(n)}$ in all dimensions. For $L^{(n)}$, the equations give
\bes{}
\bea{}\label{eq:elecL2}&&
[L^{(n)},T^{(1)}_{k}]=\Dt^{(n)}T^{(1)}_{k}+2(n+1)t^{n}T^{(1)}_{k},
\label{eq:elecL1}\\&&
[L^{(n)},T^{(2)}_{jk}]=\Dt^{(n)}T^{(2)}_{jk}+2(n+1)t^{n}T^{(2)}_{jk}
-n(n+1)t^{n-1}\big(x_{k}T^{(1)}_{j}-x_{j}T^{(1)}_{k}\big)
\nonumber\\&&\hspace{2.2cm} +\,n(n+1)t^{n-1}\big(\Delta+3-(D-1)\big)C_{jk}.
\eea\ees
When we use the equations of motion and $\Delta=\tfrac{D-2}{2}$, then the only term left is given by
\bea{}\label{eq:elecobstruction}
\big[L^{(n)},T^{(2)}_{jk}\big]\Big|_{\text{on-shell}}
=\frac{n(n+1)}{2}\,(6-D)\,t^{n-1}\big(\partial_{j}E_{k}-\partial_{k}E_{j}\big).
\eea
The factor $n(n+1)$ is zero when $n$ is $–1$ or $0$, which means that transformations in time and space, boosts, and dilatations are symmetries in every dimension. When $n$ is $1$ (in the case of the temporal SCT), and for all the higher Witt modes, the residue is proportional to ($6-D$) multiplied by the \emph{curl} of $E_i$, and disappears only if $D=6$.

\subsection{GCA invariance: magnetic limit}
\label{sec:maginv}
In the magnetic limit, we can write the equations \eqref{eq:magEOM02} as
$U_{jk}=0$ and $V_{k}=0$, where
\bea{}\label{eq:magUV}
U_{jk}=\partial^{i}H_{ijk},
\qquad
V_{k}=\partial^{i}\partial_{i}E_{k}-\partial_{k}\partial^{i}E_{i}
-\partial_{t}W_{k},
\qquad
W_{k}\equiv\partial^{i}B_{ik}.
\eea

We take \eqref{eq:Mact} and use $(a,b=1,0)$. The spatial tower ($M^{(n)}_l$) then yields
\bea{}\label{eq:magM}
[M^{(n)}_{l},U_{jk}]=-t^{n+1}\partial_{l}U_{jk},
\qquad
[M^{(n)}_{l},V_{k}]=-t^{n+1}\partial_{l}V_{k}+(n+1)t^{n}U_{kl}.
\eea
This again suggests that the equations are completely invariant under the full tower of $M$. Under the temporal tower $(L^{(n)})$, we get
\bea{}\label{eq:magL2}&&
[L^{(n)},U_{jk}]=\Dt^{(n)}U_{jk}+2(n+1)t^{n}U_{jk},
\label{eq:magL1}\\&&
[L^{(n)},V_{k}]=\Dt^{(n)}V_{k}+2(n+1)t^{n}V_{k}
-n(n+1)t^{n-1}x^{m}U_{km}
\nonumber\\&&\hspace{2.2cm}-n(n+1)t^{n-1}(\Delta-2)\,W_{k}.
\eea
Using the equations of motion and scaling weight as $\Delta-2=\tfrac{D-6}{2}$ gives
\bea{}\label{eq:magobstruction}
\big[L^{(n)},V_{k}\big]\Big|_{\text{on-shell}}
=\frac{n(n+1)}{2}\,(6-D)\,t^{n-1}\,\partial^{i}B_{ik}.
\eea
The situation is the same as in the electric case \eqref{eq:elecobstruction}, with the exception that the remaining object is now the divergence $\partial^{i}B_{ik}$. The $M^{(n)}$ tower and the global $L^{(-1,0)}$ are symmetries in all dimensions, whereas the SCT and the higher Witt modes are symmetries only when $D=6$.
\subsection{Critical dimension}
\label{sec:critdim}
Looking at the invariance of the equations in both limits under $M^{(n)}_l$ and $L^{(n)}$ tells us that they are invariant in all dimensions for all spatial tower and only $L^{(-1,0)}$. But from \eqref{eq:elecobstruction} and \eqref{eq:magobstruction}, we see that for higher Witt modes $L^{(|n|\ge1)}$, the equations are invariant only when $D=6$. This concludes that the Galilean Kalb-Ramond equations are conformally invariant if and only if 
\be{}
\boxed{\,D=6\,.}
\ee
It's the same dimension at which the original relativistic KR theory is conformal, $D_{c}=2(p+1)$ with $p=2$.
The surviving terms in \eqref{eq:elecobstruction} and \eqref{eq:magobstruction} are swapped over by the electric-magnetic map. This is stated as
\bea{}
\text{electric: curl }\;\partial_{j}E_{k}-\partial_{k}E_{j}
\qquad\longleftrightarrow\qquad
\text{magnetic: divergence }\;\partial^{i}B_{ik}.
\eea
This form of the equations points to duality that was already apparent in the representation theory as the exchange $(a,b)\rightarrow(b,a)$ of \eqref{eq:abfix-recall}. In conclusion, Galilean Kalb--Ramond theory carries the full infinite-dimensional GCA on shell at $D=6$.
\section{Null reduction and the Galilean Kalb--Ramond action}
\label{sec:nullreduction}

We have seen that the contraction of Sec(\ref{sec:contraction-invariance}) acts directly on the
equations of motion. By this method, we are not able to construct any Lagrangian for the theory. It is only applicable at the level of equations of motion. In this section, we will obtain an action for the Galilean Kalb-Ramond field by the method of null reduction ~\cite{Bagchi:2022eui}. In the null reduction method, we take a relativistic theory in $D+1$ dimensions and reduce it along the null direction. We get a larger field content, but also a local Lagrangian of a Galilean-invariant theory in $D$ dimensions. Its symmetries can be analysed off-shell. 
\subsection{Light-cone setup and reduction ansatz}
\label{sec:lc-setup}
We begin from a relativistic KR theory in $D+1$ dimensions and apply the light-cone coordinates given as
\begin{equation}
u=\tfrac{1}{\sqrt2}\left(x^1-x^0\right),\qquad t=\tfrac{1}{\sqrt2}\left(x^0+x^1\right),
\end{equation}
where the Minkowski metric takes the form 
\bea{}
ds^{2}=\eta_{\tilde\mu\tilde\nu}\,dx^{\tilde\mu}dx^{\tilde\nu}
=2\,du\,dt+\delta_{ij}\,dx^{i}dx^{j},
\qquad
\eta_{ut}=\eta_{tu}=1,\quad \eta_{ij}=\delta_{ij},
\label{eq:lcmetric-full}
\eea
with $\tilde\mu=(u,t,i)$ and $i=1,\dots,D-1$. Similarly, the inverse metric has $\eta^{ut}=\eta^{tu}=1$ and $\eta^{ij}=\delta^{ij}$, so it leaves $i$ untouched and raising an index sends $u \leftrightarrow t$. After that, we reduce along the $u$ direction. This means that all fields are taken independent of $u$ via taking $\p_u=0$. The $t$-coordinate becomes the Galilean time and the $x^i=x_i$ becomes the spatial coordinates of the Galilean theory in $ D$-dimensions. Using this reduction the two-form $B_{\tilde\mu\tilde\nu}$ decomposes into four $SO(D-1)$ multiplets,
\bea{}\label{eq:nullansatz}
\chi\equiv B_{ut}\ (\text{scalar}),\qquad
\psi_{i}\equiv B_{ui}\ (\text{vector}),\qquad
E_{i}\equiv B_{ti}\ (\text{vector}),\qquad
B_{ij}\ (\text{tensor}),
\eea
The contraction method give the multiplet $\{E_{i},B_{ij}\}$. In the null reduction method, we get two new fields $\chi$ and $\psi_{i}$ along the contraction multiplet.

\subsection{Reduced field strengths}
\label{sec:reduced-H}
The independent components of $H_{\tilde\mu\tilde\nu\tilde\lambda}=\partial_{\tilde\mu}B_{\tilde\nu\tilde\lambda}
+\partial_{\tilde\nu}B_{\tilde\lambda\tilde\mu}
+\partial_{\tilde\lambda}B_{\tilde\mu\tilde\nu}$ can be found by taking $\partial_{u}=0$ and using the identifications $B_{iu}=-\psi_{i}$, $B_{jt}=-E_{j}$. They are given by
\bes{}\label{eq:Fdef1}
\bea{}&&\label{eq:Fdef}
H_{uti}=\partial_{u}B_{ti}+\partial_{t}B_{iu}+\partial_{i}B_{ut}
=\partial_{i}\chi-\partial_{t}\psi_{i}\;\equiv\;F_{i},
\\&&\label{eq:Psidef}
H_{uij}=\partial_{u}B_{ij}+\partial_{i}B_{ju}+\partial_{j}B_{ui}
=-\big(\partial_{i}\psi_{j}-\partial_{j}\psi_{i}\big)\;\equiv\;-\Psi_{ij},
\\&&\label{eq:Gdef}
H_{tij}=\partial_{t}B_{ij}+\partial_{i}B_{jt}+\partial_{j}B_{ti}
=\partial_{t}B_{ij}-\big(\partial_{i}E_{j}-\partial_{j}E_{i}\big)\;\equiv\;G_{ij},
\\&&\label{eq:Hdef}
H_{ijk}=\partial_{i}B_{jk}+\partial_{j}B_{ki}+\partial_{k}B_{ij}.
\eea\ees
We see that the reduction of the relativistic three-form gives a
scalar-gradient
combination $F_{i}$, the curl $\Psi_{ij}$ of the null vector $\psi_{i}$, the
Galilean magnetic term $G_{ij}$, and the purely spatial three-form
$H_{ijk}$.
\subsection{The Galilean Kalb--Ramond Lagrangian}
\label{sec:GKR-lagrangian}
We will now take the null reduction of $\mathcal{L}=-\tfrac16
H_{\tilde\mu\tilde\nu\tilde\lambda}H^{\tilde\mu\tilde\nu\tilde\lambda}$. The raising of the light-cone indices ($u\leftrightarrow t$) gives
\bea{}\label{eq:raised}
H^{uti}=-F_{i},\qquad
H^{uij}=G_{ij},\qquad
H^{tij}=-\Psi_{ij},\qquad
H^{ijk}=H_{ijk}.
\eea
We know that $H$ is totally antisymmetric. Because of that,
each unordered index triple occurs $3!=6$ times in the contraction,
\bea{}
H_{\tilde\mu\tilde\nu\tilde\lambda}H^{\tilde\mu\tilde\nu\tilde\lambda}
=6\Big[\,\textstyle\sum_{i}H_{uti}H^{uti}
+\sum_{i<j}H_{uij}H^{uij}
+\sum_{i<j}H_{tij}H^{tij}
+\sum_{i<j<k}H_{ijk}H^{ijk}\Big].
\eea
Substituting \eqref{eq:Fdef1}--\eqref{eq:raised},
\bea{}
H_{uti}H^{uti}=-F_{i}F_{i},\quad
H_{uij}H^{uij}=-\Psi_{ij}G_{ij},\quad
H_{tij}H^{tij}=-\Psi_{ij}G_{ij},\quad
H_{ijk}H^{ijk}=H_{ijk}^{2},
\eea
so that, converting $2\sum_{i<j}=\sum_{ij}$ and $6\sum_{i<j<k}=\sum_{ijk}$, gives
\bea{}
H_{\tilde\mu\tilde\nu\tilde\lambda}H^{\tilde\mu\tilde\nu\tilde\lambda}=-6\,F_{i}F_{i}-6\,\Psi_{ij}G_{ij}+H_{ijk}H_{ijk}.
\eea
Therefore, the Galilean Kalb--Ramond Lagrangian is given by
\bea{}\label{eq:LGKR}
\boxed{\;
\mathcal{L}_{GKR}=F_{i}F_{i}+\Psi_{ij}G_{ij}-\tfrac16 H_{ijk}H_{ijk}\;}
\eea
or, we can write in terms of the fields,
\bea{}\label{eq:LGKRexplicit}
\mathcal{L}_{GKR}=\big(\partial_{i}\chi-\partial_{t}\psi_{i}\big)^{2}
+\big(\partial_{i}\psi_{j}-\partial_{j}\psi_{i}\big)
\big(\partial_{t}B_{ij}-\partial_{i}E_{j}+\partial_{j}E_{i}\big)
-\tfrac16 H_{ijk}H_{ijk}.
\eea
Here, the fields $\chi$ and $E_{i}$ are auxiliary and carry no time derivative, whereas $\psi_{i}$ and $B_{ij}$ are dynamical. To get the equations of motion, we will use \eqref{eq:LGKR}. They are
\bes{}\label{eq:eom1}
\begin{alignat}{2}
\delta\chi:\quad&\delta\mathcal{L}=2F_{i}\partial_{i}(\delta\chi)
&&\Rightarrow\quad \partial_{i}F_{i}=0,
\label{eq:eomchi}\\
\delta\psi_{j}:\quad&\delta\mathcal{L}=-2F_{i}\partial_{t}(\delta\psi_{i})
+2G_{ij}\partial_{i}(\delta\psi_{j})
&&\Rightarrow\quad \partial_{i}G_{ij}-\partial_{t}F_{j}=0,
\label{eq:eompsi}\\
\delta E_{j}:\quad&\delta\mathcal{L}=-2\Psi_{ij}\partial_{i}(\delta E_{j})
&&\Rightarrow\quad \partial_{i}\Psi_{ij}=0,
\label{eq:eomE}\\
\delta B_{jk}:\quad&\delta\mathcal{L}=\Psi_{jk}\partial_{t}(\delta B_{jk})
-H_{ijk}\partial_{i}(\delta B_{jk})
&&\Rightarrow\quad \partial_{i}H_{ijk}-\partial_{t}\Psi_{jk}=0.
\label{eq:eomB}
\end{alignat}
\ees
%Collecting, the Galilean Kalb--Ramond equations of motion are
%\begin{equation}
%\partial_{i}F_{i}=0,\qquad
%\partial_{i}G_{ij}-\partial_{t}F_{j}=0,\qquad
%\partial_{i}\Psi_{ij}=0,\qquad
%\partial_{i}H_{ijk}-\partial_{t}\Psi_{jk}=0.
%\label{eq:GKReom}
%\end{equation}
To reduce to the magnetic sector, we switch off two auxiliary null fields, $\chi=\psi_{i}=0$, the remaining equations are
\bea{}\label{eq:mag-from-null}
\partial_{i}G_{ij}=0
\;\Longrightarrow\;
\partial_{i}\partial_{i}E_{j}-\partial_{j}\partial_{i}E_{i}
-\partial_{t}\partial_{i}B_{ij}=0,
\qquad
\partial_{i}H_{ijk}=0,
\eea
These are the magnetic limit equations that were obtained in Sec.~\ref{sec:contraction-invariance}.
\subsection{Invariance under GCA}
\label{sec:GKR-action-inv}
Before looking at the invariance of \eqref{eq:LGKR} and \eqref{eq:eom1}, we will see the scaling weights for all fields present in \eqref{eq:LGKR}. We use the anisotropic scaling $t\to\lambda^{z}t$, $x^{i}\to\lambda x^{i}$. The invariance of
$\int dt\,d^{D-1}x\,\mathcal{L}_{GKR}$ needs every term to carry weight $(D+z-1)$. We demand that the terms in \eqref{eq:Fdef1} scale homogeneously fixes the weight for all terms $(F_{i}, G_{ij}, F^{2}, \Psi G, H^{2})$ and the conditions comes out to be
\bea{}\label{eq:GKR-dims}
\Delta_{\chi}=\Delta_{B}=\frac{z+D-3}{2},\qquad
\Delta_{\psi}=\frac{D-1-z}{2},\qquad
\Delta_{E}=\frac{3z+D-5}{2}.
\eea
%through $F_{i}$ ($\Delta_{\chi}+1=\Delta_{\psi}+z$) and $G_{ij}$
For the Galilean conformal case, we take $z=1$, and the scaling dimensions for each field become
\bea{}\label{eq:GKR-commondim}
\Delta_{\chi}=\Delta_{\psi}=\Delta_{E}=\Delta_{B}=\frac{D-2}{2},
\eea
The weights (5.13) make the action scale-invariant for any $z$. As such, the free theory does not specify the value of the dynamical exponent, making $z=1$ a freely chosen parameter, motivated by the GCA. This is due to the fact that two independent scalings of the parent theory remain after reduction: relativistic dilatation and boost $J_{ut}$ in the reduced null direction. This contrasts with the Galilean Yang--Mills theory, where fixing $z+D=5$ is enforced by the cubic coupling \cite{Bagchi:2022eui}. In our case, however, we have a free abelian theory, which means only adding an interaction (such as a Chapline--Manton type coupling $H=dB-\kappa\,A\wedge F$ to a one-form) would be able to specify $z$. We will leave this for future work.

Using them, we can see that the action \eqref{eq:LGKR} is invariant under scale transformation. But we can also see it by using the action of the scale transformation seen in the representation theory Sec(\ref{sec:rep}). We use \eqref{eq:Mact} and $\Delta=\tfrac{D-2}{2}$, then each field strength in \eqref{eq:Fdef1} has weight $\Delta+1=\tfrac{D}{2}$, so the Lagrangian \eqref{eq:LGKR} becomes
\bea{}
\delta_{D}\mathcal{L}_{GKR}=\big(t\partial_{t}+x^{l}\partial_{l}+D\big)\mathcal{L}_{GKR}
=\partial_{t}\big(t\mathcal{L}_{GKR}\big)+\partial_{l}\big(x^{l}\mathcal{L}_{GKR}\big),
\eea
a total derivative.
Moving on to the Galilean Boost $B_{k}=M^{(0)}_{k}$, the field strengths transforms as
\bes{}\label{boosts}
\bea{}&&
\delta_{B}F_{i}=-t\partial_{k}F_{i}-\Psi_{ik},
\delta_{B}\Psi_{ij}=-t\partial_{k}\Psi_{ij},
\label{eq:boostFPsi}\\&&
\delta_{B}G_{ij}=-t\partial_{k}G_{ij}-H_{kij}-\delta_{ik}F_{j}+\delta_{jk}F_{i},\\&&
\delta_{B}H_{ijl}=-t\partial_{k}H_{ijl}
+\delta_{jk}\Psi_{il}+\delta_{lk}\Psi_{ji}+\delta_{ik}\Psi_{lj}.
\label{eq:boostGH}
\eea\ees
Using them for the variation of \eqref{eq:LGKR}, the only term survive is 
\bea{}\label{eq:boost-tot-der}
\delta_{B}\mathcal{L}_{GKR}=-t\partial_{k}\mathcal{L}_{GKR}=-\partial_{k}\big(t\mathcal{L}_{GKR}\big),
\eea
again, a total derivative.
%\begin{equation}
%S_{k}[\chi]=-\psi_{k},\quad
%S_{k}[\psi_{i}]=0,\quad
%S_{k}[E_{i}]=B_{ik}-\delta_{ik}\chi,\quad
%S_{k}[B_{ij}]=\delta_{ik}\psi_{j}-\delta_{jk}\psi_{i},
%\label{eq:Sk-recall}
%\end{equation}
%together with the field action \eqref{eq:Lact}--\eqref{eq:Mact} of the infinite
%GCA.
The action is therefore invariant under dilatation and
under boosts. We will now move forward and see how the action \eqref{eq:LGKR} vary under the infinite symmetries $(M^{(n)}_{l}, L^{(n)})$ of GCA. The variation of action \eqref{eq:LGKR} under spatial part of infinite GCA \eqref{eq:Mact} gives 
\bea{}\label{eq:Mn-obstruction}
\delta_{M^{(n)}_{l}}\mathcal{L}_{GKR}
=-\partial_{l}\big(t^{n+1}\mathcal{L}_{GKR}\big)+2\,n(n+1)\,t^{n-1}\,\Psi_{lj}\psi_{j}.
\eea
The $M^{(n)}_{l}$ part of infinite GCA is a symmetry for $n=-1,0$ but not for $|n|\ge1$. Similarly, we will look for the variation of the action \eqref{eq:LGKR} under the temporal part of infinite GCA. The remaining terms that survive are given by
%\paragraph{Field strengths under $M^{(n)}_{l}$.}
%Using $[M^{(n)}_{l},\Phi]=-t^{n+1}\partial_{l}\Phi+(n+1)t^{n}S_{l}[\Phi]$ with the
%boost data \eqref{eq:Sk-recall}, the field strengths transform as
%\begin{align}
%\delta_{M^{(n)}_{l}}F_{i}&=-t^{n+1}\partial_{l}F_{i}-(n+1)t^{n}\Psi_{il},\\
%\delta_{M^{(n)}_{l}}\Psi_{ij}&=-t^{n+1}\partial_{l}\Psi_{ij},\\
%\delta_{M^{(n)}_{l}}G_{ij}&=-t^{n+1}\partial_{l}G_{ij}
%-(n+1)t^{n}\big(H_{lij}+\delta_{il}F_{j}-\delta_{jl}F_{i}\big)
%+n(n+1)t^{n-1}\big(\delta_{il}\psi_{j}-\delta_{jl}\psi_{i}\big),\\
%\delta_{M^{(n)}_{l}}H_{ijk}&=-t^{n+1}\partial_{l}H_{ijk}
%+(n+1)t^{n}\big(\delta_{jl}\Psi_{ik}-\delta_{kl}\Psi_{ij}-\delta_{il}\Psi_{jk}\big),
%\end{align}
%reducing to \eqref{eq:boostFPsi}--\eqref{eq:boostGH} at $n=0$.
%\paragraph{Variation of $\LGKR$ under $M^{(n)}_{l}$.}
\bea{}\label{eq:Ln-obstruction}&&
\delta_{L^{(n)}}\mathcal{L}_{GKR}
=\partial_{t}\big(t^{n+1}\mathcal{L}_{GKR}\big)+(n+1)\,\partial_{l}\big(t^{n}x^{l}\mathcal{L}_{GKR}\big)
\nonumber\\&&\hspace{2.2cm}+\,n(n+1)t^{n-1}\Big[2(1-\Delta)\,F_{i}\psi_{i}
+(\Delta-2)\,\Psi_{ij}B_{ij}\Big]\nonumber\\&&\hspace{2.2cm}
\;-\;2\,n(n+1)(n-1)t^{n-2}\,x_{i}\psi_{j}\Psi_{ij}.
\eea
Here also the $L^{(n)}$ part of infinite GCA is a symmetry for $n=-1,0$ but not for $|n|\ge1$.

We first looked at the invariance of \eqref{eq:LGKRexplicit} under GCA, now we will see what happens to the equations of motion \eqref{eq:eom1}. For simplicity, we will write the equations as
\bea{}\label{eq:Edef}
E^{(1)}=\partial_{i}F_{i},~
E^{(2)}_{j}=\partial_{i}G_{ij}-\partial_{t}F_{j},~
E^{(3)}_{j}=\partial_{i}\Psi_{ij},~
E^{(4)}_{jl}=\partial_{i}H_{ijl}-\partial_{t}\Psi_{jl}.
\eea
The $M^{(n)}_{l}$ generators act on \eqref{eq:Edef} to give
%Acting with the boost and using the cyclic identity
%$\partial_{i}H_{kij}=\partial_{i}H_{ijk}$, every equation maps into a combination
%of equations,
%\begin{alignat}{2}
%\delta_{B}E^{(1)}&=-t\partial_{k}E^{(1)}-E^{(3)}_{k},&\qquad
%\delta_{B}E^{(3)}_{j}&=-t\partial_{k}E^{(3)}_{j},
%\label{eq:boostE13}\\
%\delta_{B}E^{(2)}_{j}&=-t\partial_{k}E^{(2)}_{j}-E^{(4)}_{jk}+\delta_{jk}E^{(1)},&\qquad
%\delta_{B}E^{(4)}_{jl}&=-t\partial_{k}E^{(4)}_{jl}
%+\delta_{jk}E^{(3)}_{l}-\delta_{lk}E^{(3)}_{j}.
%\label{eq:boostE24}
%\end{alignat}
%The system is therefore boost invariant, and the same closure holds for all
%$M^{(n)}_{k}$ and for the finite $L^{(-1,0,1)}$. With the auxiliary fields
%switched off ($\chi=\psi=0$, so $F_{i}=\Psi_{ij}=0$) the system collapses to the
%magnetic-sector pair $E^{(2)}_{j}=E^{(4)}_{jl}=0$, which by
%Sec.~\ref{sec:contraction-invariance} enjoys the entire infinite-dimensional GCA
%at $D=6$.
\bes{}
\bea{}&&
\delta_{M^{(n)}_{m}}E^{(1)}=-t^{n+1}\partial_{m}E^{(1)}-(n+1)t^{n}E^{(3)}_{m},\\&&
\delta_{M^{(n)}_{m}}E^{(3)}_{j}=-t^{n+1}\partial_{m}E^{(3)}_{j},\\&&
\delta_{M^{(n)}_{m}}E^{(2)}_{j}=-t^{n+1}\partial_{m}E^{(2)}_{j}
-(n+1)t^{n}E^{(4)}_{jm}+(n+1)t^{n}\delta_{jm}E^{(1)}
\nonumber\\&&\hspace{2.2cm}+n(n+1)t^{n-1}\big(\partial_{j}\psi_{m}-\delta_{jm}\partial_{i}\psi_{i}\big),\\&&
\delta_{M^{(n)}_{m}}E^{(4)}_{jk}=-t^{n+1}\partial_{m}E^{(4)}_{jk}
+(n+1)t^{n}\big(\delta_{jm}E^{(3)}_{k}-\delta_{km}E^{(3)}_{j}\big),
\eea\ees
Similarly, under the temporal tower. we have 
\bes{}\bea{}\label{eq:Ln-E4}&&\hspace{-.5cm}
\delta_{L^{(n)}}E^{(1)}=\Dt^{(n)}_{\Delta+2}E^{(1)}
+n(n+1)t^{n-1}\Big[(1-\Delta)\partial_{i}\psi_{i}+x^{k}E^{(3)}_{k}\Big],\\&&\hspace{-.5cm}
\delta_{L^{(n)}}E^{(3)}_{j}=\Dt^{(n)}_{\Delta+2}E^{(3)}_{j},\\&&\hspace{-.5cm}
\delta_{L^{(n)}}E^{(2)}_{j}=\Dt^{(n)}_{\Delta+2}E^{(2)}_{j}
+n(n+1)t^{n-1}\Big[x^{k}E^{(4)}_{jk}-x_{j}E^{(1)}
+(\Delta-2)\partial^{i}B_{ij}+\tfrac{D-4}{2}F_{j}-(1-\Delta)\partial_{t}\psi_{j}\Big]
\nonumber\\&&\hspace{-.5cm}
\hspace{2.2cm}-\,n(n+1)(n-1)t^{n-2}\Big[\tfrac{D}{2}\psi_{j}
+x^{k}\partial_{k}\psi_{j}-x_{j}\partial_{i}\psi_{i}+x^{k}\Psi_{jk}\Big],\\&&\hspace{-.5cm}
\delta_{L^{(n)}}E^{(4)}_{jk}=\Dt^{(n)}_{\Delta+2}E^{(4)}_{jk}
+n(n+1)t^{n-1}\Big[x_{k}E^{(3)}_{j}-x_{j}E^{(3)}_{k}
+\tfrac{D-6}{2}\Psi_{jk}\Big].
\eea\ees
where $\Dt^{(n)}\equiv t^{n+1}\partial_{t}+(n+1)t^{n}(x^{l}\partial_{l}+\Delta)$.
Here, we also see the same structure showing up as seen when we look at the variation of action \eqref{eq:LGKRexplicit} under the infinite section of GCA. These equations are invariant under $M^{(n)}_{l}$ and $L^{(n)}$ for $n=-1,0$ but not for $|n|\ge1$.

\subsection{Origin of the obstruction: comparison with Galilean electrodynamics}
We will repeat the analysis for the null reduction of Maxwell theory, where we have
\bea{}
\mathcal L_{\rm GED}=\frac12(\partial_t\phi)^2+E_i\partial_i\phi-\frac14W_{ij}W_{ij},\eea 
where $E_i =(\p_t a_i - \p_i a_t)$ and $W_{ij}=(\p_i a_j - \p_j a_i)$. Here, $(a_t,a_i,\phi)$ are the Galilean gauge and scalar fields.  We use the same representation \eqref{symm}, with boost data $S_k[\phi]=0$, $S_k[a_t]=-a_k$, $S_k[a_i]=\delta_{ik}\phi$ and $\Delta=\frac{D-2}{2}$. We get
\begin{equation}
\delta_{M^{(n)}_l}\mathcal L_{\rm GED}=\partial_l\Big[-t^{n+1}\mathcal L_{\rm GED}+\tfrac{n(n+1)}{2}t^{n-1}\phi^2\Big],
\end{equation}
so the whole $M^{(n)}$ tower is a symmetry in any $D$. The temporal generators $L^{(-1,0,1)}$ are symmetries precisely in $D=4$ \cite{Bagchi:2022eui}. For the two-form the corresponding term in \eqref{eq:Mn-obstruction} is
\begin{equation}
2n(n+1)\,t^{n-1}\,\Psi_{lj}\psi_j=n(n+1)\,t^{n-1}\Big[\partial_l(\psi_j\psi_j)-2\,\psi_j\partial_j\psi_l\Big],
\end{equation}
and the last term is not a total derivative. The difference comes from the boost chain lowest component ($S_k[\cdot]=0$). In the case of $p=1$ this is given by the scalar $A_u$; in the case of $p=2$ it is instead given by the vector $\psi_i=B_{ui}$, whose antisymmetrised derivative $\Psi_{ij}$ appears in the action. The lowest component for a $ p$-form should be the $(p-1)$ form $B_{u i_1\dots i_{p-1}}$, and hence the obstruction persists for all $p\ge2$. However, we can still see the effects of the critical dimension in the null reduction. When $\Delta=2$, the term $(\Delta -2)\Psi_{ij} B_{ij}$ in \eqref{eq:Ln-obstruction} is zero; hence all obstructions in \eqref{eq:Mn-obstruction}-\eqref{eq:Ln-obstruction} are proportional to $\psi_i$. Hence, if we set $\chi=\psi_i=0$, i.e. truncate to the magnetic sector, then the symmetry obtained by contraction is recovered. Finally, note that $K, K_i$ do not come from the conformal group of the parent theory. If they are symmetries of a null reduced action such as GED, then they are emergent symmetries.

\subsection{Schr\"odinger invariance in \texorpdfstring{$D=5$}{D=5}}
\label{sec:schrodinger}
Some of the symmetry properties of the reduced action \eqref{eq:LGKR} are inherited from the relativistic theory. A conformal Killing vector of $(D+1)$-dimensional Minkowski space transforms $u$-independent configurations into $u$-independent ones if and only if its commutator with $\partial_u$ is proportional to $\partial_u$. Of the generators of $\mathfrak{so}(D+1,2)$ this picks out
\bea{}\label{eq:preserved}
  P_t,\quad P_i,\quad P_u,\quad J_{ij},\quad J_{iu},\quad J_{ut},\quad \tilde D,\quad \tilde K_u ,
\eea
while $J_{it}$, $\tilde K_t$ and $\tilde K_i$ are lost. In the reduced theory the
central element $P_u$ acts as zero and the $J_{iu}$ are the Galilean boosts. The generators \eqref{eq:preserved} span the Schr\"odinger algebra
$\mathfrak{sch}(D-1)$ with mass $P_u$
\cite{Hagen:1972pd,Niederer:1972zz,Son:2008ye,Balasubramanian:2008dm,Duval:2008jg},
extended by the additional dilatation $J_{ut}$. The Galilean conformal generators
$K=L^{(1)}$ and $K_i=M^{(1)}_i$ are not in \eqref{eq:preserved}; when they are
symmetries of a null-reduced action, as for Galilean electrodynamics (GED) in
four dimensions~\cite{Bagchi:2022twx}, they are emergent rather than inherited.

This also leads to the Schr\"odinger invariance of the three-dimensional GED obtained by null reducing the critical 4D Maxwell theory \cite{Chapman:2020vtn,Bagchi:2022twx}.
The Kalb--Ramond analogue is the reduction of the critical six-dimensional theory, i.e.\ $D=5$, not $D=6$; at $D=6$ the GCA arises instead from the contraction of Sec.\eqref{sec:contraction}.
We elaborate on the correspondence below and point out where it fails.

\paragraph{Dilatations and scaling weights:}
The combination of $\tilde D$ and $J_{ut}$ generates
$t\to\lambda^z t$, $x^i\to\lambda x^i$, $u\to\lambda^{2-z}u$ for every $z$.
Since the integral over $u$ is not part of the reduced action, the rescaling of
$u$ is compensated by a shift $-(2-z)/2$ of the field weights. A component with
$n_t$ lower $t$-indices and $n_u$ lower $u$-indices then has the reduced weight
\bea{}\label{eq:weights-general}
  \Delta_\Phi \;=\; \tilde\Delta-\frac{2-z}{2}+(z-1)\,(n_t-n_u),
  \qquad \tilde\Delta=\frac{D-1}{2},
\eea
where $\tilde\Delta$ is the canonical dimension \eqref{eq:canonical} of the parent field.
For the two-form, \eqref{eq:weights-general} reproduces \eqref{eq:GKR-dims}. Applied to
the one-form, $(A_u,A_t,A_i)=(\phi,a_t,a_i)$, it gives
\bea{}\label{eq:weights-GED}
  \Delta_\phi=\frac{D-1-z}{2},\qquad
  \Delta_{a_i}=\frac{D-3+z}{2},\qquad
  \Delta_{a_t}=\frac{D-5+3z}{2},
\eea
which equals $(0,1,2)$ at $D=3$, $z=2$ and $(1,1,1)$ at $D=4$, $z=1$, in
agreement with \cite{Bagchi:2022twx}. In both cases, the
null component ($A_u$, respectively $\psi_i=B_{ui}$) sits at the bottom of the
boost chain and carries the lowest weight. Since \eqref{eq:weights-general} holds for all $z$, two independent scalings of the parent theory survive in the free theory; thus, the dynamical exponent is unfixed by the free theory. The compensating shift disappears at $z=2$, as it should since $\tilde K_u$ does not scale $u$.
\bea{}\label{eq:weights-z2}
  (\Delta_\chi,\Delta_\psi,\Delta_E,\Delta_B)=(2,1,3,2),
\eea
and the $z=2$ dilatation acts as
$\delta\Phi=(2t\partial_t+x^k\partial_k+\Delta_\Phi)\Phi$.

\paragraph{Special conformal transformation:}
On $u$-independent fields $\tilde K_u$ acts through
\bea{}\label{eq:xiC}
  \xi_C=t^2\partial_t+t\,x^i\partial_i-\tfrac12 x^ix^i\partial_u,
  \qquad \mathcal L_{\xi_C}\eta=2t\,\eta ,
\eea
and the two-form transforms as a primary of dimension $\tilde\Delta$,
\bea{}\label{eq:deltaC-B}
  \delta_C B=\mathcal L_{\xi_C}B+(\tilde\Delta-2)\,t\,B .
\eea
The $\partial_u$ component of $\xi_C$ drops out on $u$-independent fields, but
contributes to the Lie derivative through $\partial_i\xi^u=-x_i$.\footnote{%
$\xi_C$ is one half of the generator $\tilde K$ of~\cite{Bagchi:2022twx},
eq.~(2.9). Our light-cone coordinate $u$ also differs in sign from the one used
there, so the transformations below have been rederived in the present
conventions rather than transcribed.}
Evaluating \eqref{eq:deltaC-B} component by component gives, for every $D$,
\bea{}\label{eq:deltaC-compact}
  \delta_C\Phi=\big(t^2\partial_t+t\,x^k\partial_k+t\,\Delta_\Phi\big)\Phi-x^k S_k[\Phi],
\eea
where $\Delta_\Phi$ are the $z=2$ weights \eqref{eq:weights-general} and
$S_k$ is the boost data \eqref{eq:nullboost}. The Weyl term $(\tilde\Delta-2)t=\frac{D-5}{2}\,t$
is what converts the geometric weights of the Lie derivative into the $z=2$
weights. In components, for $D=5$,
\bes{}\label{eq:deltaC-components}
\bea{}&&
  \delta_C\chi=\big(t^2\partial_t+t\,x^k\partial_k+2t\big)\chi+x^k\psi_k, ~~
  \delta_C\psi_i=\big(t^2\partial_t+t\,x^k\partial_k+t\big)\psi_i,\\&&
  \delta_C E_i=\big(t^2\partial_t+t\,x^k\partial_k+3t\big)E_i+x^kB_{ki}+x_i\chi,\\&& 
  \delta_C B_{ij}=\big(t^2\partial_t+t\,x^k\partial_k+2t\big)B_{ij}-x_i\psi_j+x_j\psi_i .
\eea
\ees
These are the two-form counterparts given in \cite{Bagchi:2022twx}. The
same construction applied to the one-form gives
$\delta_C a_t=(\cdots+2t)a_t+x^ka_k$, $\delta_C a_i=(\cdots+t)a_i-x_i\phi$ and no
algebraic term for $\phi$. Like $\phi$, the vector $\psi_i$ transforms purely
orbitally. In the conventions of \eqref{symm}, $[C,\Phi]=-\delta_C\Phi$.
Using \eqref{eq:LGKR}, the Lagrangian transforms as
\bea{}\label{eq:deltaC-L}
  \delta_C\mathcal L_{\rm GKR}
  =\partial_t\big(t^2\mathcal L_{\rm GKR}\big)+\partial_i\big(t\,x^i\mathcal L_{\rm GKR}\big)
  +\frac{D-5}{2}\sum_\Phi\Pi_\Phi\,\Phi ,
\eea
with $\Pi_\Phi=\partial\mathcal L_{\rm GKR}/\partial(\partial_t\Phi)$, so that $\sum_\Phi\Pi_\Phi\,\Phi=-2F_i\psi_i+\Psi_{ij}B_{ij}$. The last
term is not a total derivative. It is the reduced form of the relativistic
obstruction \eqref{eq:dHfull} and vanishes precisely when the parent theory is
critical. Together with translations, rotations and boosts, this establishes:
\begin{quote}
\emph{Null reduction of six-dimensional conformal Kalb--Ramond theory results in a five-dimensional Galilean theory with an action which is invariant under the Schr\"odinger algebra with $z=2$ extended by the dilatation $J_{ut}$. As we have seen previously, the null reduction of four-dimensional Maxwell theory results in Schr\"odinger invariant GED in 3D~\cite{Chapman:2020vtn, Bagchi:2022twx}.}
\end{quote}

\paragraph{The Schr\"odinger-Virasoro extension:}
For three-dimensional GED, more than the finite algebra is realised: the action
is also invariant under the full tower of Schr\"odinger boosts $Y^{(m)}_i$, and
the equations with $\phi=0$ are invariant under the complete
Schr\"odinger--Virasoro algebra~\cite{Bagchi:2022twx,Henkel:1993sg}. To test the
corresponding statements for the two-form, we define, for arbitrary functions
$f(t)$ and $g(t)$,
\bes{}\label{eq:SV-action}
\bea{}&&
  \delta_{Y_f}\Phi=-f\,\partial_l\Phi+f'\,S_l[\Phi],\\&&
  \delta_{Z_g}\Phi=\Big[g\,\partial_t+\tfrac12 g'\big(x^k\partial_k+\Delta_\Phi\big)\Big]\Phi
  -\tfrac12 g''\,x^kS_k[\Phi],
\eea
\ees
with the $z=2$ weights. These are the $z=2$ analogues of \oldeq{3.27}. The modes
$f=t^{m+1/2}$, $m\in\mathbb Z+\tfrac12$, and $g=t^{n+1}$, $n\in\mathbb Z$, give
the generators $Y^{(m)}_l$ and $Z^{(n)}$ of~\cite{Bagchi:2022twx}. In particular $f=1,t$ give $P_l$ and the boost
$G_l$, while $g=1,t,t^2$ give $H$, $\tfrac12\tilde D$ and $\xi_C$ of
\eqref{eq:deltaC-compact}. For arbitrary $f$ and $g$ one finds
\bea{}&&
  \delta_{Y_f}\mathcal L_{\rm GKR}=-\partial_l\big(f\,\mathcal L_{\rm GKR}\big)
  +2f''\,\Psi_{lj}\psi_j, \label{eq:dY-L}\\&&
  \delta_{Z_g}\mathcal L_{\rm GKR}=\partial_t\big(g\,\mathcal L_{\rm GKR}\big)
  +\partial_k\big(\tfrac12 g'x^k\mathcal L_{\rm GKR}\big)
  -g'''\,x_i\psi_j\Psi_{ij}
  +\frac{D-5}{4}\,g''\sum_\Phi\Pi_\Phi\,\Phi . \label{eq:dZ-L}
\eea
For $f=t^{n+1}$, \eqref{eq:dY-L} reduces to \eqref{eq:Mn-obstruction}: the spatial tower does
not involve the weights and its obstruction is independent of $z$. Neither
$\Psi_{lj}\psi_j$ nor $x_i\psi_j\Psi_{ij}$ is a total derivative; for instance,
\begin{equation}\label{eq:EL-obstruction}
  \frac{\delta}{\delta\psi_m}\int\Psi_{lj}\psi_j=\delta_{lm}\,\partial_k\psi_k-\partial_m\psi_l\neq0 .
\end{equation}
In $D=5$ this implies the following:
\begin{itemize}
\item \emph{Schr\"odinger boosts:} For $f=t^{m+1/2}$ one has
  $f''=(m^2-\tfrac14)\,t^{m-3/2}$, so only $m=\pm\tfrac12$, i.e.\ $P_l$ and
  $G_l$, are symmetries of the action. 
The higher $Y^{(m)}_l$ are not, and hence, since they do not vanish on shell, they are also not symmetries of the equations of motion. This is where the analogy to the GED fails. In the GED we find the same computation yielding the obstruction $f''\phi\,\partial_l\phi=\partial_l(\tfrac12 f''\phi^2)$, which is a total derivative and therefore all $Y^{(m)}_i$ are symmetries of the three-dimensional GED action \cite{Bagchi:2022twx}.
\item \emph{Higher temporal modes:} For $g=t^{n+1}$,
  $g'''=(n+1)n(n-1)\,t^{n-2}$, so only $Z^{(-1,0,1)}=(H,\tfrac12\tilde D,\xi_C)$
  survive. This is as in GED, where the corresponding obstruction is proportional
  to $n(n^2-1)\,t^{n-2}\phi^2$~\cite{Bagchi:2022twx}; for the two-form it is
  again built from $\psi_i$.
\end{itemize}

\paragraph{The magnetic truncation:}
At $D=5$ every obstruction in \eqref{eq:dY-L}--\eqref{eq:dZ-L} is built from
$\psi_i$. The truncation $\chi=\psi_i=0$ is consistent with
\eqref{eq:SV-action}, since $\delta\psi_i\propto\psi$ and $\delta\chi\propto(\chi,\psi)$,
and it leaves the magnetic equations \eqref{eq:mag-from-null},
$\mathcal G_j\equiv\partial_iG_{ij}=0$ and
$\mathcal H_{jk}\equiv\partial_iH_{ijk}=0$. With
$\mathcal O_g\equiv g\,\partial_t+\tfrac12 g'x^k\partial_k$, they transform as
\bea{}&&
  \delta_{Y_f}\mathcal G_j=-f\,\partial_l\mathcal G_j+f'\,\mathcal H_{lj},\qquad
  \delta_{Y_f}\mathcal H_{jk}=-f\,\partial_l\mathcal H_{jk}, \label{eq:trunc-Y}\\&&
  \delta_{Z_g}\mathcal G_j=\Big[\mathcal O_g+\tfrac12 g'(\Delta_B+3)\Big]\mathcal G_j
  -\tfrac12 g''\,x^k\mathcal H_{kj}
  +\tfrac12 g''\,(\Delta_B-2)\,\partial_iB_{ij},\nonumber\\&&
  \delta_{Z_g}\mathcal H_{jk}=\Big[\mathcal O_g+\tfrac12 g'(\Delta_B+2)\Big]\mathcal H_{jk}.
  \label{eq:trunc-Z}
\eea
Since $\Delta_B-2=\frac{D-5}{2}$ at $z=2$, the magnetic equations in $D=5$ are
covariant under the full Schr\"odinger--Virasoro algebra, for arbitrary $f$ and
$g$. This is the $z=2$ counterpart of \oldeq{4.13}, with the critical dimension
now entering as $D+1=6$. The pattern is the same as in three-dimensional
GED~\cite{Bagchi:2022twx}, except that for the two-form the infinite tower of
Schr\"odinger boosts is lost as soon as the null fields are kept.
Table~\ref{tab:sch-summary} summarises the comparison.

\begin{table}[ht]
\centering
\renewcommand{\arraystretch}{1.25}
\begin{tabular}{lcc}
\toprule
 & GED, $D=3$~\cite{Bagchi:2022twx} & Kalb--Ramond, $D=5$ \\
\midrule
Finite Schr\"odinger algebra, action & \checkmark & \checkmark \\
$Y^{(m)}_i$, $|m|\geq\tfrac32$, action and equations & \checkmark & $\times$ \\
$Z^{(n)}$, $|n|\geq 2$, action and full equations & $\times$ & $\times$ \\
Full Schr\"odinger--Virasoro, truncated equations & \checkmark\ ($\phi=0$) & \checkmark\ ($\chi=\psi_i=0$) \\
\bottomrule
\end{tabular}
\caption{Schr\"odinger symmetry of null-reduced gauge theories whose parent
theory is critical.}
\label{tab:sch-summary}
\end{table}

For GED, the modes $Z^{(n)}$ that fail as symmetries of the action are
nevertheless realised as Hamiltonian vector fields on phase space, with a
state-dependent central extension~\cite{Bagchi:2022twx}. Whether the same happens
for $Z^{(n)}$, and for the higher $Y^{(m)}_i$, in the Kalb--Ramond theory is a
question for the canonical analysis.

\subsection{Sectors with non-zero null momentum}\label{snznm}
We know that $\partial_u$ is a Killing vector. The parent field can be expanded in eigenmodes 
\bea{}
B_{\tilde\mu\tilde\nu}=e^{imu}\,b_{\tilde\mu\tilde\nu}(t,x)+{\rm c.c.}.
\eea
The components of $b$ are given as a set $(\chi,\psi_i,E_i,B_{ij})$. We repeat the Secs[\eqref{sec:reduced-H}, \eqref{sec:GKR-lagrangian}] with $\partial_u\to im$ gives 
\bea{} H_{uti}=F_i+imE_i, H_{uij}=-(\Psi_{ij}-imB_{ij}), H_{tij}=G_{ij}\eea and, finally we get
\bea{}
\mathcal L_m=\big|F_i+imE_i\big|^2+\tfrac12\Big[\overline{(\Psi_{ij}-imB_{ij})}\,G_{ij}+{\rm c.c.}\Big]-\tfrac16\,\bar H_{ijk}H_{ijk},
\eea
which reduces to \eqref{eq:LGKR} at $m=0$. For $m\neq0$ a gauge transformation with parameter $\theta_{\tilde\mu}\propto e^{imu}$, under which $\delta\psi_i=im\theta_i-\partial_i\theta_u$ and $\delta\chi=im\theta_t-\partial_t\theta_u$, can be used to impose the light-cone gauge $\chi=\psi_i=0$. The residual parameters $\theta_{\tilde\mu}=\partial_{\tilde\mu}(\theta_u/im)$ are the gauge-for-gauge symmetry and act trivially. $E_i$ is then auxiliary, with solution $E_j=\frac{i}{m}\partial_iB_{ij}$. The $\chi$- and $\psi_i$-equations are implied by the remaining equations of motion. Substituting, and dropping total derivatives,
\begin{equation}\label{eq:schrodinger-B}
\mathcal L_m=\sum_{i<j}\Big[\,im\big(\bar B_{ij}\partial_tB_{ij}-B_{ij}\partial_t\bar B_{ij}\big)-\partial_k\bar B_{ij}\,\partial_kB_{ij}\Big],\qquad
i\partial_tB_{ij}=-\frac{1}{2m}\nabla^2B_{ij} .
\end{equation}
The non-zero-momentum sectors are therefore $\frac{(D-1)(D-2)}{2}$ free Schr\"odinger fields of mass $m$ in the antisymmetric representation of $SO(D-1)$. Their number equals the degree-of-freedom count (2.11) of the parent theory, as it must for a light-cone description. Their propagator,
\begin{equation}
\langle B_{ij}(\omega,\mathbf k)\,\bar B_{kl}(\omega,\mathbf k)\rangle=\frac{i\,T_{ijkl}}{2m\omega-\mathbf k^2+i\epsilon},
\end{equation}
fixes $z=2$ dynamically, in contrast with the zero-momentum sector (Sec.~6.3). For $D+1=6$ there are six fields, which split under $SO(4)$ into self-dual and anti-self-dual parts, $3+3$. The chiral two-form of the six-dimensional $(2,0)$ tensor multiplet keeps only one of these triplets.
\section{Correlation function}
\label{sec:corre-function}
We will now see what happens to the correlation functions in the Galilean limit of Kalb-Ramond theory. Till now, we have seen that how we get the Galilean limit from both ways, namely, the contraction method and the null reduction method. We have next looked at the invariance of the theory under the finite and infinite parts of GCA. For the correlation function, we will first see the Ward identities related to finite GCA and check the forms in both methods. 

There are two comments that can be made here. Firstly, note that $E_i$ and $B_{ij}$ (or $\chi$ and $\psi_i$ after null reduction) are the components of the gauge potential. Like the relativistic correlator \eqref{eq:BB2pt}, the two-point functions of the gauge potential are gauge-dependent, and so it is an assumption on the kinematics to consider these objects as Galilean conformal primaries. The expressions we have derived below are the most general possible two-point functions consistent with the global Ward identities, and not any specific gauge-fixed theory. However, the field strengths do contain gauge-invariant information; namely, the components of $H$. Secondly, the free null-reduced theory realises one particular solution of these identities. This gauge-fixed propagator is frequency-independent, and so all free field correlators are supported on $\tau=0$. These solutions to the identities are not captured by our power law ansatz. Furthermore, the boost and special conformal identities are trivially satisfied. Therefore, the power laws we have derived should be interpreted as the most general structure available to a Galilean conformal theory with this field content. Since the contraction of Sec.~4 does not possess an action, but does have equations of motion, these are all we can know. In this section, we will use the notation given below: points $1=(t_1,x_1)$, $2=(t_2,x_2)$;
\be{}
\tau\equiv t_1-t_2,\qquad x\equiv x_1-x_2,\qquad r\equiv x\!\cdot\!x=|x_1-x_2|^2,
\ee
and short form of correlators as $\corr{\Phi_A}{\Phi_B}\equiv\langle0|\Phi_A(t_1,x_1)\Phi_B(t_2,x_2)|0\rangle$, always in
this order.
\paragraph{Translations and Rotations:}
We will take the general form of the generator as $\circledast$. As per the property, they annihilate the vacuum on both sides,
$\circledast|0\rangle=0=\langle0|\circledast$,
\be{}
0=\langle0|[\circledast,\Phi_A(1)\Phi_B(2)]|0\rangle
=\corr{[\circledast,\Phi_A(1)]}{\Phi_B(2)}+\corr{\Phi_A(1)}{[\circledast,\Phi_B(2)]}.
\label{eq:master}
\ee
The correlation functions under space and time translations are given by the condition
\be{}
(\p_{t_1}+\p_{t_2})G=0, ~(\p_{1,l}+\p_{2,l})G=0 \implies G=G(\tau,x)
\ee
where we have used $[H,\Phi]=\p_t\Phi$ and $[P_l,\Phi]=-\p_l\Phi$ with $G\equiv\corr{\Phi_A}{\Phi_B}$. Next, rotation ($J_{ij}$) require $G$ to be an $SO(D-1)$-covariant tensor. It is made out of $\delta_{ij}$, $x_i$, and scalars $f(\tau,r)$.
\paragraph{Scale Transformation:}
We use $[\bar{D},\Phi]=(t\p_t+x^l\p_l+\Delta)\Phi$, so that the constraint equations comes out to be
\be{}\label{syf}
t_1\p_{t_1}G+t_2\p_{t_2}G=(t_1-t_2)\,\p_\tau G=\tau\p_\tau G,\qquad
x_1^l\p_{1,l}G+x_2^l\p_{2,l}G=(x_1-x_2)^l\p_lG=x^l\p_lG.
\ee
Now, we use $\p_{t_1}G=\p_\tau G$, $\p_{t_2}G=-\p_\tau G$ and similarly for space, \eqref{syf} becomes the Euler equation
\be{}\label{eq:dil}
\big(\tau\p_\tau+x^l\p_l+2\Delta\big)G=0
\qquad\Longleftrightarrow\qquad
G(\lambda\tau,\lambda x)=\lambda^{-2\Delta}G(\tau,x),
\ee
Nothing else is allowed except $x^{(p)}\,\tau^{-2\Delta-p}$ (a $p$-th degree monomial in $x$ times
the compensating power of $\tau$).
\paragraph{Boost and Special Conformal Transformations:}
For them, we will use the spatial part of Infinite GCA, which is $[M^{(n)}_l,\Phi(t,x)]=-t^{n+1}\p_l\Phi+(n+1)t^nS_l[\Phi]$ at $n=0$ (boost $B_l$) and $n=1$ (spatial SCT $K_l$). We also take $t_1\p_{1,l}+t_2\p_{2,l}=(t_1-t_2)\p_l=\tau\p_l$ and
$t_1^2\p_{1,l}+t_2^2\p_{2,l}=(t_1^2-t_2^2)\p_l=\tau(t_1+t_2)\p_l$ on $G(\tau,x)$. Together, while applying to \eqref{eq:master}, they give
\bea{}\label{eq:B}
\tau\,\p_l\corr{\Phi_A}{\Phi_B} =\corr{S_l[\Phi_A]}{\Phi_B}+\corr{\Phi_A}{S_l[\Phi_B]}
\eea
\bea{}
\tau(t_1+t_2)\,\p_l\corr{\Phi_A}{\Phi_B} =2t_1\corr{S_l[\Phi_A]}{\Phi_B}+2t_2\corr{\Phi_A}{S_l[\Phi_B]}.\label{eq:K}
\eea
\paragraph{Infinite GCA:}
We have seen above the Ward identities for translations, rotations, boosts, scale and special conformal transformations. But we don't know what to expect from an infinite tower of GCA. Since $L^{(n)}, M^{(n)}$ are the symmetries of the equations of motion or action at the critical dimension, but a Ward identity needs the vacuum to be annihilated on both sides, which holds only for $n\in\{-1,0,1\}$. To have a look into it, let us apply the $L^{(n)}$ identity to an $x$-independent power law $G=\tau^{-2\Delta}$. The identity then becomes $\big[t_1^{n+1}\p_{t_1}+t_2^{n+1}\p_{t_2}+(n+1)(t_1^n+t_2^n)\Delta\big]G=0$, with $\p_{t_1}G=-2\Delta\tau^{-2\Delta-1}$, $\p_{t_2}G=+2\Delta\tau^{-2\Delta-1}$,
\bea{}\label{eq:tower}
\frac{\text{residual}}{G}
=-2\Delta\,\frac{t_1^{n+1}-t_2^{n+1}}{t_1-t_2}+(n+1)\Delta\,(t_1^n+t_2^n)
\nonumber\\\hspace{-1cm}=\begin{cases}
0, & n=-1,\\
-2\Delta+2\Delta=0, & n=0,\\
-2\Delta(t_1{+}t_2)+2\Delta(t_1{+}t_2)=0, & n=1,\\
\nonumber
\Delta\,(t_1-t_2)^2, & n=2,\\
2\Delta\,(t_1-t_2)^2(t_1+t_2), & n=3.
\end{cases}\\
\eea
As seen from above, even the correlator of a perfectly good primary violates the naive $n\ge2$ identity. In contrast to the $L^{(n)}$ tower, the $M^{(n)}$-tower is satisfied for \emph{all} $n$ by the surviving correlators.
\paragraph{Two lemmas:}
We have constructed these two lemmas from the properties of the Ward identities. These lemmas will be very beneficial for finding the form of all correlation functions from the contraction method as well as the null reduction method. They are given below: 
\begin{lemma}[bottom of the chain]\label{lem:1} For both entries, if  $S_l[\Phi]=0$, then \eqref{eq:B} tells $\tau\p_l G=0$, so $G$ is $x$-independent; the rotation covariance pushes the unique $\delta$-saturation of the free indices (or $G=0$ if none exists), and \eqref{eq:dil} sets the power:
$G=C\,(\text{$\delta$'s})\,\tau^{-2\Delta}$.
\end{lemma}
%\begin{lemma}[bottom of the chain]\label{lem:1}
%If $S_l[\Phi]=0$ for both entries, \eqref{eq:B} reads $\tau\p_l G=0$, so $G$ is
%$x$-independent; rotation covariance forces the unique $\delta$-saturation of the free
%indices (or $G=0$ if none exists), and \eqref{eq:dil} fixes the power:
%$G=C\,(\text{$\delta$'s})\,\tau^{-2\Delta}$.
%\end{lemma}
\begin{lemma}[integration constants]\label{lem:2}
We get an ambiguity that is an $x$-independent covariant tensor of the same free-index type when we integrate $\p_lG=(\text{source})_l$. We have found that the only $x$-independent invariant tensor is $\delta_{ij}$ (parity-even theory). We can get a new constant iff the free indices of $G$ can be completely saturated by $\delta$'s respecting the index symmetries. Iff the total number of valences is even and no antisymmetric pair must be built from $\delta$'s alone. The correlators with an odd number of valences integrate uniquely.  
\end{lemma}
%\begin{lemma}[integration constants]\label{lem:2}
%When integrating $\p_lG=(\text{source})_l$, the ambiguity is an $x$-independent covariant
%tensor of the same free-index type. Since the only $x$-independent invariant tensor is
%$\delta_{ij}$ (parity-even theory), a new constant enters iff the free indices of $G$ can be
%completely saturated by $\delta$'s respecting the index symmetries --- i.e.\ iff the total
%valence is even and no antisymmetric pair must be built from $\delta$'s alone. Odd-valence
%correlators integrate uniquely.
%\end{lemma}

\subsection{Correlation functions via contraction method}
We will look into the correlation functions in both limits, namely, the electric and magnetic limit of Galilean Kalb-Ramond theory. For that, we will take $\Delta=\frac{D-2}{2}=2$, fields $\{E_i,B_{ij}\}$. From the representation theory, the boost acts on the fields at $t=0, x=0$ as
\bea{}
S_l[E_m]=a\,B_{ml},\qquad S_l[B_{mn}]=b\,(\delta_{ln}E_m-\delta_{lm}E_n),\qquad ab=0 .
\eea
\paragraph{Electric limit:}
\paragraph{Step 1: $\corr{E_{i}}{E_{j}}$:} From the representation theory, the boost acting on $E_{i}$ is zero. So, we have $S_l[E]=0$ on both entries. This means that Lemma~\ref{lem:1} applies. Rotation acting on 2-index $x$-independent tensor leaves only $\delta_{ij}$. Under dilatation, we finally get
\bea{}\label{eq:elEE}
\corr{E_i}{E_j}=C_E\,\delta_{ij}\,\tau^{-4}.
\eea
\paragraph{Step 2: $\corr{B_{ij}}{E_k}$ and $\corr{{E_i} B_{kl}}$:}
We know that $S_l[B_{ij}]=\delta_{lj}E_i-\delta_{li}E_j$ and $S_l[E_k]=0$. Under boost \eqref{eq:B}, we get 
\bea{}
\tau\,\p_l\corr{B_{ij}}{E_k}
=\delta_{lj}\corr{E_i}{E_k}-\delta_{li}\corr{E_j}{E_k}
=C_E\,\tau^{-4}\big(\delta_{lj}\delta_{ik}-\delta_{li}\delta_{jk}\big).
\eea
From above, we have $\p_l\corr{B_{ij}}{E_k}=C_E\tau^{-5}(\delta_{lj}\delta_{ik}-\delta_{li}\delta_{jk})$. So, taking an integration on both side gives
\be{}\label{eq:elBE}
\corr{B_{ij}}{E_k}=C_E\,\tau^{-5}\big(x_j\delta_{ik}-x_i\delta_{jk}\big).
\ee
We see that there is no integration constant. This is because the homogeneous piece would be an $x$-independent 3-index tensor. So, the Lemma~\ref{lem:2} applies. Similarly, we get
\be{}
\corr{E_i}{B_{kl}}=C_E\,\tau^{-5}\big(x_l\delta_{ik}-x_k\delta_{il}\big).
\label{eq:elEB}
\ee
\paragraph{Step 3: $\corr{B_{ij}}{B_{kl}}$:} Under the boost transformation, we get
\bea{}\label{eq:elBBsource}
\tau\,\p_m\corr{B_{ij}}{B_{kl}}
&=\delta_{mj}\corr{E_i}{B_{kl}}-\delta_{mi}\corr{E_j}{B_{kl}}
 +\delta_{ml}\corr{B_{ij}}{E_k}-\delta_{mk}\corr{B_{ij}}{E_l}\notag\\[2pt]
&=C_E\tau^{-5}\Big[\delta_{mj}(x_l\delta_{ik}-x_k\delta_{il})
 -\delta_{mi}(x_l\delta_{jk}-x_k\delta_{jl})\notag\\
&\hspace{16mm}+\delta_{ml}(x_j\delta_{ik}-x_i\delta_{jk})
 -\delta_{mk}(x_j\delta_{il}-x_i\delta_{jl})\Big].
\eea
For the sake of simplicity, we define two invariant structures
\bea{}
T_{ijkl}\equiv\delta_{ik}\delta_{jl}-\delta_{il}\delta_{jk},\qquad
Q_{ijkl}\equiv x_ix_k\delta_{jl}-x_ix_l\delta_{jk}-x_jx_k\delta_{il}+x_jx_l\delta_{ik}.
\eea
We can also check how $Q_{ijkl}$ is related to the term in \eqref{eq:elBBsource}. For that, we differentiate $Q$ and get
\be{}
\p_mQ_{ijkl}
=\delta_{mi}\,(x_k\delta_{jl}-x_l\delta_{jk})
+\delta_{mj}\,(x_l\delta_{ik}-x_k\delta_{il})
+\delta_{mk}\,(x_i\delta_{jl}-x_j\delta_{il})
+\delta_{ml}\,(x_j\delta_{ik}-x_i\delta_{jk}),
\label{eq:dQ}
\ee
As we can see, they are the same term. Hence, the result becomes $C_E\tau^{-6}Q$. We also have a homogeneous ambiguity and it is given as  $C_B\, T\,\tau^{-4}$. The complete solution takes the form as
\be{}\label{eq:elBB}
\boxed{\;\corr{B_{ij}}{B_{kl}}=C_B\,\tau^{-4}\,T_{ijkl}+C_E\,\tau^{-6}\,Q_{ijkl}\;}
\qquad
\ee
Equations \eqref{eq:elEE}--\eqref{eq:elBB} are the complete solution of the boost hierarchy. We will now see what happens to these correlations when we impose a special conformal transformation. 
For that, we will apply \eqref{eq:K} to $(E_i(1),B_{kl}(2))$, we get
\bea{}
\text{LHS}&=\tau(t_1+t_2)\,\p_m\corr{E_i}{B_{kl}}
=C_E\,(t_1+t_2)\,\tau^{-4}\big(\delta_{ml}\delta_{ik}-\delta_{mk}\delta_{il}\big),\\
\text{RHS}&=2t_2\big[\delta_{ml}\corr{E_i}{E_k}-\delta_{mk}\corr{E_i}{E_l}\big]
=2t_2\,C_E\,\tau^{-4}\big(\delta_{ml}\delta_{ik}-\delta_{mk}\delta_{il}\big).
\eea
The difference comes out to be 
\be{}
\text{LHS}-\text{RHS}
=C_E\,(t_1+t_2-2t_2)\,\tau^{-4}(\cdots)
=C_E\,(t_1-t_2)\,\tau^{-4}(\cdots)
=C_E\,\tau^{-3}\big(\delta_{ml}\delta_{ik}-\delta_{mk}\delta_{il}\big),
\ee
which must vanish for independent $t_1,t_2$:
\be{}
\boxed{\,C_E=0\,}\qquad\text{(electric sector).}
\ee
We see that $C_E=0$, so the second term in \eqref{eq:elBB} disappears. Similarly, \eqref{eq:elEE}--\eqref{eq:elEB} vanishes. The final form of all correlators in the electric limit is given by
\be{}
\boxed{\;\corr{B_{ij}}{B_{kl}}=C_B\,(\delta_{ik}\delta_{jl}-\delta_{il}\delta_{jk})\,
(t_1-t_2)^{-4},\qquad
\corr{E_i}{E_j}=\corr{E_i}{B_{kl}}=\corr{B_{ij}}{E_k}=0.}
\ee
\paragraph{Magnetic limit:}
In the magnetic limit, the boost acts on the fields at $t=0,x=0$ in opposite to the one in electric limit. Here, we have $S_l[E_i]=B_{il}$, $S_l[B]=0$.

\paragraph{Step 1: $\corr{B_{ij}}{B_{kl}}$:}
We will use Lemma~\ref{lem:1}. The invariant that respects the antisymmetry of both pairs is $T_{ijkl}$. It is x-independent and has a degree $-4$. We get this when we act on the correlator with translation, rotation, dilatation and boost. The correlator becomes
\be{}
\corr{B_{ij}}{B_{kl}}=C_B\,\tau^{-4}\,T_{ijkl}.
\label{eq:magBB}
\ee
\paragraph{Step 2: $\corr{E_i}{B_{kl}}$ and $\corr{B_{ij}}{E_k}$:} Under Boost, this becomes
\be{}
\tau\,\p_m\corr{E_i}{B_{kl}}=\corr{B_{im}}{B_{kl}}
=C_B\tau^{-4}\big(\delta_{ik}\delta_{ml}-\delta_{il}\delta_{mk}\big)
\;\Longrightarrow\;
\corr{E_i}{B_{kl}}=C_B\,\tau^{-5}\big(x_l\delta_{ik}-x_k\delta_{il}\big),
\ee
Similarly,
\be{}
\tau\,\p_m\corr{B_{ij}}{E_k}=\corr{B_{ij}}{B_{km}}
=C_B\tau^{-4}\big(\delta_{ik}\delta_{jm}-\delta_{im}\delta_{jk}\big)
\;\Longrightarrow\;
\corr{B_{ij}}{E_k}=C_B\,\tau^{-5}\big(\delta_{ik}x_j-x_i\delta_{jk}\big).
\ee
\paragraph{Step 3: $\corr{E_i}{E_j}$:} Same as above, we will first apply the boost. We get 
\bea{}\label{eq:magEEsource}&&
\tau\,\p_l\corr{E_i}{E_j}
=\corr{B_{il}}{E_j}+\corr{E_i}{B_{jl}}\notag\\&&
=C_B\tau^{-5}\big[(\delta_{ij}x_l-x_i\delta_{lj})+(x_l\delta_{ij}-x_j\delta_{il})\big]
=C_B\tau^{-5}\big[2x_l\delta_{ij}-x_i\delta_{lj}-x_j\delta_{il}\big].
\eea
To make it more simplified, we can use $\p_l(r\delta_{ij})=2x_l\delta_{ij}$ and $\p_l(x_ix_j)=\delta_{li}x_j+x_i\delta_{lj}$, therefore, the solution becomes
$C_B\tau^{-6}(r\delta_{ij}-x_ix_j)$. Apart from this, we also get a homogeneous piece $C_E\delta_{ij}\tau^{-4}$. The complete result is
\be{}\label{eq:magEE}
\boxed{\;\corr{E_i}{E_j}=C_E\,\delta_{ij}\,\tau^{-4}
+C_B\,\tau^{-6}\big(r\,\delta_{ij}-x_ix_j\big)\;}
\ee
Now, we will impose $K$ and see how the correlations form changes. Take $(E_i(1),E_j(2))$, and apply $K$, we get
\bea{}&&
\text{LHS}=\tau(t_1+t_2)\,\p_l\corr{E_i}{E_j}
=C_B(t_1+t_2)\,\tau^{-5}\big[2x_l\delta_{ij}-x_i\delta_{lj}-x_j\delta_{il}\big],\\&&
\text{RHS}=2t_1\corr{B_{il}}{E_j}+2t_2\corr{E_i}{B_{jl}}
=2t_1C_B\tau^{-5}(\delta_{ij}x_l-x_i\delta_{lj})+2t_2C_B\tau^{-5}(x_l\delta_{ij}-x_j\delta_{il}).\nonumber
\eea
We have three independent structures. We will collect the coefficients. They are
\be{}
x_l\delta_{ij}:\;2(t_1{+}t_2)-2t_1-2t_2=0,~
x_i\delta_{lj}:\;-(t_1{+}t_2)+2t_1=t_1{-}t_2,~
x_j\delta_{il}:\;-(t_1{+}t_2)+2t_2=t_2{-}t_1,
\ee
so, we get the (LHS-RHS) as
\bea{}
C_B\,(t_1-t_2)\,\tau^{-5}\big(x_i\delta_{lj}-x_j\delta_{il}\big)
=C_B\,\tau^{-4}\big(x_i\delta_{lj}-x_j\delta_{il}\big)
\;\Longrightarrow\;\boxed{\,C_B=0\,.}
\eea
The correlation functions in the magnetic limit become
\be{}
\boxed{\;\corr{E_i}{E_j}=C_E\,\delta_{ij}\,(t_1-t_2)^{-4},\qquad
\corr{B_{ij}}{B_{kl}}=\corr{E_i}{B_{kl}}=\corr{B_{ij}}{E_k}=0.\;}
\ee
In two limits, we can see some duality. The two \emph{(B)-families} are exchanged by $E\leftrightarrow B$, $C_E\leftrightarrow C_B$. When we impose $K$, it deletes the constant belonging to the boost-\emph{inert} field, and with it every cross-correlator.
%\subsection{Duality, sharpened}
%The two \emph{(B)-families} are already exchanged by $E\leftrightarrow B$,
%$C_E\leftrightarrow C_B$: compare \eqref{eq:elEE}--\eqref{eq:elBB} with
%\eqref{eq:magBB}--\eqref{eq:magEE}. Imposing (K) then deletes, in each sector, exactly the
%constant belonging to the boost-\emph{inert} field, and with it every tail and every cross
%correlator. Away from $D=6$, (K) is unavailable and the two-parameter families with tails
%are the final answer.
\subsection{Correlation functions via Null-reduction method}
We will look into the correlation functions in the null-reduction method. For that, we will take $\Delta=\frac{D-2}{2}=2$, fields $\{\chi,\psi_i,E_i,B_{ij}\}$. From the representation theory, the boost acts on the fields at $t=0, x=0$ as
\be{}\label{eq:Sdata}
S_k[\chi]=-\psi_k,\qquad S_k[\psi_i]=0,\qquad
S_k[E_i]=B_{ik}-\delta_{ik}\chi,\qquad
S_k[B_{ij}]=\delta_{ik}\psi_j-\delta_{jk}\psi_i .
\ee
The map $S$ is nilpotent with the filtration
\be{}
E\;\xrightarrow{\;S\;}\;\{B,\chi\}\;\xrightarrow{\;S\;}\;\psi\;\xrightarrow{\;S\;}\;0.
\ee
The fields are graded as $g(\psi)=0$, $g(\chi)=g(B)=1$, $g(E)=2$. We then solve pairs in order of
total grade $g_A+g_B$.
%so we grade the fields $g(\psi)=0$, $g(\chi)=g(B)=1$, $g(E)=2$ and solve pairs in order of
%total grade $g_A+g_B$: each Ward identity sources a correlator only through correlators of
%strictly lower total grade, already known.
\paragraph{Grade 0: $\corr{\psi_i}{\psi_j}$:} 
\begin{itemize}
    \item $\psi_i-\psi_j$ Correlator: The action of translations and rotations on \eqref{eq:master} allows $\corr{\psi_i}{\psi_j}=f_1(\tau,r)\delta_{ij}+f_2(\tau,r)x_ix_j$. Since
 $S[\psi]=0$, \eqref{eq:B} gives 
 \bea{}&&
\tau\p_l\corr{\psi_i}{\psi_j}= 2x_lf_1'\delta_{ij}+(2x_lf_2'x_ix_j+f_2(\delta_{li}x_j+x_i\delta_{lj}))=0
 \eea
Here, primes are derivatives with respect to $r$. Boost transformation forces $f_2=0$, $f_1'=0$. Further, scale transformation \eqref{eq:dil} fixes the correlation function as
\be{}\label{eq:pp}
\boxed{\;\corr{\psi_i}{\psi_j}=C_\psi\,\delta_{ij}\,\tau^{-2\Delta}.\;}
\ee
\end{itemize}
\paragraph{Grade 1: $\corr{\chi}{\psi_j}$, $\corr{B_{ij}}{\psi_k}$:} 
\begin{itemize}
    \item $\chi-\psi_j$ Correlator: The correlation function $\corr{\chi}{\psi_j}$ under boost transformation becomes
\bea{}
\tau\p_k\corr{\chi}{\psi_j}=-\corr{\psi_k}{\psi_j} \equiv -C_\psi\delta_{kj}\tau^{-2\Delta}
\eea
Now taking $\tau$ on the RHS and integrating on both sides to get
\bea{}\label{eq:cp}
\corr{\chi}{\psi_j}=-C_\psi\,x_j\,\tp{1}.
\eea
Similarly for $\corr{\psi_j}{\chi}$, we have the same as \eqref{eq:cp}.
\item $B_{ij}-\psi_k$ Correlator: Now, we look at $\corr{B_{ij}}{\psi_k}$. Under a boost transformation, the correlation function becomes
\bea{}
\tau\p_m\corr{B_{ij}}{\psi_k}=\delta_{im}\corr{\psi_j}{\psi_k}-\delta_{jm}\corr{\psi_i}{\psi_k}\equiv C_\psi\tau^{-2\Delta}(\delta_{im}\delta_{jk}-\delta_{jm}\delta_{ik})
\eea
Next, we take the integration on both sides and get
\be{}\label{eq:Bp}
\corr{B_{ij}}{\psi_k}=\corr{\psi_k}{B_{ij}}
=C_\psi\,\tp{1}\big(x_i\delta_{jk}-x_j\delta_{ik}\big).
\ee
\end{itemize}

\paragraph{Grade 2: $\corr{\chi}{\chi}$, $\corr{B_{ij}}{\chi}$, $\corr{B_{ij}}{B_{kl}}$, $\corr{E_i}{\psi_j}$:}
For all cases, we will see what happens to the correlators when we apply it with boost transformations. They are given as
\begin{itemize}
    \item $\chi$-$\chi$ Correlator: \bea{}
\tau\p_k\corr{\chi}{\chi}=-\corr{\psi_k}{\chi}-\corr{\chi}{\psi_k}
=2C_\psi x_k\tp{1}\nonumber\\
\;\Longrightarrow\;
\p_k\corr{\chi}{\chi}=2C_\psi x_k\tp{2}=C_\psi\,\p_k(r)\,\tp{2}.
\eea
We see that the count for the free index is zero (even), so a constant is allowed. The scale transformation \eqref{eq:dil} fixes the weight power as
\be{}\label{eq:cc}
\boxed{\;\corr{\chi}{\chi}=C_\chi\,\tau^{-2\Delta}+C_\psi\,r\,\tp{2}.\;}
\ee
\item $B_{ij}$--$\chi$ Correlator: Using boost transformation, we get 
\bea{}\label{eq:Bchicancel}
\tau\p_m\corr{B_{ij}}{\chi}
=\delta_{im}\corr{\psi_j}{\chi}-\delta_{jm}\corr{\psi_i}{\chi}
-\corr{B_{ij}}{\psi_m}=0
%\notag\\
%=-C_\psi\tp{1}\big[\underbrace{\delta_{im}x_j-\delta_{jm}x_i}_{\text{from }S[B]}=0
%+\underbrace{x_i\delta_{jm}-x_j\delta_{im}}_{\text{from }S[\chi]}\big]=0 :
\eea
Hence, this correlation function is x-independent. We have
\be{}\label{eq:Bc}
\boxed{\;\corr{B_{ij}}{\chi}=\corr{\chi}{B_{ij}}=0.\;}
\ee
\item $B_{ij}$--$B_{kl}$ Correlator: Under the Galilean transformations, we have
\begin{align*}
\tau\p_m\corr{B_{ij}}{B_{kl}}
&=\delta_{im}\corr{\psi_j}{B_{kl}}-\delta_{jm}\corr{\psi_i}{B_{kl}}
+\delta_{km}\corr{B_{ij}}{\psi_l}-\delta_{lm}\corr{B_{ij}}{\psi_k}\\
&=C_\psi\tp{1}\Big[\delta_{im}(x_k\delta_{jl}-x_l\delta_{jk})
-\delta_{jm}(x_k\delta_{il}-x_l\delta_{ik})\\
&\hspace{18mm}
+\delta_{km}(x_i\delta_{jl}-x_j\delta_{il})
-\delta_{lm}(x_i\delta_{jk}-x_j\delta_{ik})\Big],
\end{align*}
Similar to the contraction case, $rT$ is excluded and $T$ enters freely:
\bea{}\label{eq:BB}
\boxed{\;\corr{B_{ij}}{B_{kl}}=C_B\,\tau^{-2\Delta}\,T_{ijkl}+C_\psi\,\tp{2}\,Q_{ijkl}.\;}
\eea
\item $E_i-\psi_j$ Correlator: Under boost transformation, we have
\bea{}\hspace{-.5cm}
\tau\p_l\corr{E_i}{\psi_j}
=\corr{B_{il}}{\psi_j}-\delta_{il}\corr{\chi}{\psi_j}
=C_\psi\tp{1}\big(x_i\delta_{lj}-x_l\delta_{ij}\big)+C_\psi\,\delta_{il}\,x_j\tp{1},
\eea
It also admits a constant, since we have valence of $2$. The complete solution becomes
%so $\p_l\corr{E_i}{\psi_j}=C_\psi\tp{2}\,(x_i\delta_{lj}+x_j\delta_{il}-x_l\delta_{ij})$.
%Since
%\[
%%\p_l\Big(x_ix_j-\tfrac{r}{2}\delta_{ij}\Big)
%=\delta_{li}x_j+x_i\delta_{lj}-x_l\delta_{ij},
%\]
\bea{}\label{eq:Ep}
\corr{E_i}{\psi_j}=C_\psi\,\tp{2}\Big(x_ix_j-\tfrac{r}{2}\delta_{ij}\Big)
+C_1\,\delta_{ij}\,\tau^{-2\Delta}.
\eea
The correlation function $\corr{\psi_j}{E_i}$ also give the same result but with a new independent constant $C'_{1}$. Each order is constrained separately by the Ward identities and does not relate $C_1$ and $C_1'$. On the other hand, the Hermiticity of the fields tells $C_1^{*}=C_1'$.

\bea{}
\corr{\psi_j}{E_i}=C_\psi\,\tp{2}\Big(x_ix_j-\tfrac{r}{2}\delta_{ij}\Big)
+C_1'\,\delta_{ij}\,\tau^{-2\Delta}.
\label{eq:pE}
\eea
\end{itemize}
%The reversed ordering satisfies the \emph{same} differential equation
%($\tau\p_l\corr{\psi_j}{E_i}=\corr{\psi_j}{B_{il}}-\delta_{il}\corr{\psi_j}{\chi}$, same
%right-hand side), but its integration constant is independent:

%Unlike the grade-1 correlators, nothing forces $C_1=C_1'$: the Ward identities never relate
%opposite operator orderings, and a relation (e.g.\ from Hermiticity and a spectral condition)
%would be extra input beyond the algebra.
\paragraph{Grade 3: $\corr{E_i}{\chi}$ and $\corr{E_i}{B_{kl}}$:}
\begin{itemize}
    \item $E_i-\chi$ Correlator: Similar to above, we will look under boost transformations. The correlation function becomes
    \bea{}&&\nonumber
\tau\p_l\corr{E_i}{\chi}
=\corr{B_{il}}{\chi}-\delta_{il}\corr{\chi}{\chi}-\corr{E_i}{\psi_l}\\\nonumber&&
=0-\delta_{il}\big(C_\chi\tau^{-2\Delta}+C_\psi r\tp{2}\big)
-C_\psi\tp{2}\Big(x_ix_l-\tfrac{r}{2}\delta_{il}\Big)-C_1\delta_{il}\tau^{-2\Delta}\\&&
=-(C_\chi+C_1)\,\delta_{il}\,\tau^{-2\Delta}
-C_\psi\,\tp{2}\Big(x_ix_l+\tfrac{r}{2}\delta_{il}\Big).
\eea
Taking the integral, we get
\bea{}\label{eq:Ec}
\corr{E_i}{\chi}=-(C_\chi+C_1)\,x_i\,\tp{1}-\tfrac{C_\psi}{2}\,x_i\,r\,\tp{3},
\eea
%begin{align*}
%\p_l f_i&=-(C_\chi+C_1)\delta_{il}\tp{1}
%-\tfrac{C_\psi}{2}\big(\delta_{il}r+2x_ix_l\big)\tp{3}\\
%&=-(C_\chi+C_1)\delta_{il}\tp{1}-C_\psi\tp{3}\Big(x_ix_l+\tfrac{r}{2}\delta_{il}\Big).
%\;\checkmark
%\end{align*}
%Valence $1$ is odd, so no constant:
    \item $E_i-B_{kl}$ Correlator: Under boost, we have
    \bea{}\label{eq:EBraw}&&
\tau\p_m\corr{E_i}{B_{kl}}
=\corr{B_{im}}{B_{kl}}-\delta_{im}\underbrace{\corr{\chi}{B_{kl}}}_{=0}
+\delta_{km}\corr{E_i}{\psi_l}-\delta_{lm}\corr{E_i}{\psi_k}\notag\\&&
=C_B\tau^{-2\Delta}\big(\delta_{ik}\delta_{ml}-\delta_{il}\delta_{mk}\big)
+C_\psi\tp{2}\big(x_ix_k\delta_{ml}-x_ix_l\delta_{mk}-x_mx_k\delta_{il}+x_mx_l\delta_{ik}\big)
\notag\\&&
\quad+C_\psi\tp{2}\big(\delta_{km}x_ix_l-\delta_{lm}x_ix_k\big)
-\tfrac{C_\psi}{2}r\tp{2}\,R_{m}
+C_1\tau^{-2\Delta}\,R_{m}.
\eea
where $R_{m;i,kl}\equiv\delta_{km}\delta_{il}-\delta_{lm}\delta_{ik}$. 
After doing all calculations and computations, we get
\bea{}\label{eq:EBclean}
\tau\p_m\corr{E_i}{B_{kl}}
=(C_1-C_B)\,\tau^{-2\Delta}\,R_m
-C_\psi\,\tp{2}\Big(x_m\,P+\tfrac{r}{2}\,R_m\Big),
\eea
with $P_{i;kl}\equiv x_k\delta_{il}-x_l\delta_{ik}$. Doing the integration, the result becomes
\bea{}\label{eq:EB}
\corr{E_i}{B_{kl}}=(C_1-C_B)\,\tp{1}\,\big(x_k\delta_{il}-x_l\delta_{ik}\big)
-\tfrac{C_\psi}{2}\,\tp{3}\,r\,\big(x_k\delta_{il}-x_l\delta_{ik}\big),
\eea
Similarly, for $\corr{B_{kl}{E_i}}$, we get the same expression with $C_1$ exchanged with $C_{1}^{'}$.
\end{itemize}
%Two cancellations now occur. First, using $\delta_{ml}=\delta_{lm}$, $\delta_{mk}=\delta_{km}$,
%\begin{equation}
%x_ix_k\delta_{ml}-\delta_{lm}x_ix_k=0,\qquad
%-\,x_ix_l\delta_{mk}+\delta_{km}x_ix_l=0 :
%%\label{eq:magic}
%\end{equation}
%the structures $x_ix_k\delta_{ml}$, which are \emph{not} the $\p_m$ of any covariant tensor
%antisymmetric in $(k,l)$ (the only candidates $P$ and $rP$ produce $R_m$, $x_mP$ and $rR_m$
%under $\p_m$, nothing else), drop out identically between the $\corr{B}{B}$ tail and the
%$\corr{E}{\psi}$ tail. This is the integrability condition
%$\p_n(\text{source})_m=\p_m(\text{source})_n$ holding by virtue of the Jacobi identity of
%the representation --- had any coefficient in \eqref{eq:Sdata} been different, the hierarchy
%would have been inconsistent here. Second, the $\delta\delta$ terms merge:
%$\delta_{ik}\delta_{ml}=\delta_{lm}\delta_{ik}$ and $\delta_{il}\delta_{mk}=\delta_{km}\delta_{il}$
%give $C_B(\delta_{ik}\delta_{ml}-\delta_{il}\delta_{mk})=-C_B R_m$, hence
%using $-x_mx_k\delta_{il}+x_mx_l\delta_{ik}=-x_mP$. The general covariant ansatz is
%$\corr{E_i}{B_{kl}}=a\,\tp{1}P+b\,\tp{3}rP$ (the would-be third structure
%$x_ix_kx_l$ dies under antisymmetrization in $(k,l)$, and $x_i\delta_{kl}$ dies by symmetry
%of $\delta$), with
%\[
%\p_m\big(a\,\tp{1}P+b\,\tp{3}rP\big)
%=a\,\tp{1}R_m+b\,\tp{3}\big(2x_mP+rR_m\big).
%\]
%Matching \eqref{eq:EBclean}/$\tau$: from $x_mP$, $2b=-C_\psi$; from $rR_m$, $b=-C_\psi/2$
%(consistent --- one equation more than unknowns, satisfied); from $R_m$, $a=C_1-C_B$.
%Valence $3$: no constant.
\paragraph{Grade 4: $\corr{E_i}{E_j}$}
\begin{itemize}
    \item $E_i - E_j$ Correlator: In the grade 4 category, we only have one correlation function. Under boost, we have
    \bea{}\label{eq:EEward}
\tau\p_l\corr{E_i}{E_j}
=\corr{B_{il}}{E_j}-\delta_{il}\corr{\chi}{E_j}
+\corr{E_i}{B_{jl}}-\delta_{jl}\corr{E_i}{\chi}.
\eea
Using all the correlation functions from all grades, we end up with
\bea{}&&
\p_l\corr{E_i}{E_j}
=(C_1{+}C_1'{+}C_\chi{-}C_B)\tp{2}\,\p_l(x_ix_j)
\nonumber\\&&\hspace{2.2cm}+\Big(C_B{-}\tfrac{C_1+C_1'}{2}\Big)\tp{2}\,\p_l(r)\,\delta_{ij}
+\tfrac{C_\psi}{4}\tp{4}\,\p_l(r^2)\,\delta_{ij},
\eea
Solving the integral, we have
\bea{}
\boxed{\;\begin{aligned}
\corr{E_i}{E_j}
&=C_E\,\delta_{ij}\,\tau^{-2\Delta}
+(C_1{+}C_1'{+}C_\chi{-}C_B)\,x_ix_j\,\tp{2}\\
&\quad+\Big(C_B-\tfrac{C_1+C_1'}{2}\Big)\,r\,\delta_{ij}\,\tp{2}
+\tfrac{C_\psi}{4}\,r^2\,\delta_{ij}\,\tp{4}.
\end{aligned}\;}
\label{eq:EE}
\eea
%Insert \eqref{eq:Ec}--\eqref{eq:BE} with the appropriate index substitutions:
%\begin{align*}
%\corr{B_{il}}{E_j}&=(C_1'-C_B)\tp{1}(x_i\delta_{lj}-x_l\delta_{ij})
%-\tfrac{C_\psi}{2}\tp{3}r\,(x_i\delta_{lj}-x_l\delta_{ij}),\\
%\corr{E_i}{B_{jl}}&=(C_1-C_B)\tp{1}(x_j\delta_{il}-x_l\delta_{ij})
%-\tfrac{C_\psi}{2}\tp{3}r\,(x_j\delta_{il}-x_l\delta_{ij}),\\
%-\delta_{il}\corr{\chi}{E_j}&=+(C_\chi+C_1')\,\delta_{il}x_j\tp{1}
%\tfrac{C_\psi}{2}\,\delta_{il}\,x_j\,r\,\tp{3},\\
%\delta_{jl}\corr{E_i}{\chi}&=+(C_\chi+C_1)\,\delta_{jl}x_i\tp{1}
%+\tfrac{C_\psi}{2}\,\delta_{jl}\,x_i\,r\,\tp{3}.
%\end{align*}
%Collect the $\tp{1}$ terms by structure:
%%\begin{align*}
%%x_i\delta_{lj}&:\quad (C_1'-C_B)+(C_\chi+C_1)=C_1+C_1'+C_\chi-C_B,\\
%x_j\delta_{il}&:\quad (C_1-C_B)+(C_\chi+C_1')=C_1+C_1'+C_\chi-C_B,\\
%x_l\delta_{ij}&:\quad -(C_1'-C_B)-(C_1-C_B)=2C_B-C_1-C_1' ,
%\end{align*}
%and the $\tp{3}$ terms:
%\[
%-\tfrac{C_\psi}{2}r\big[x_i\delta_{lj}+x_j\delta_{il}-2x_l\delta_{ij}\big]
%+\tfrac{C_\psi}{2}r\big[\delta_{il}x_j+\delta_{jl}x_i\big]
%=C_\psi\,r\,x_l\,\delta_{ij}
%\]
%--- the $x_i\delta_{lj}$ and $x_j\delta_{il}$ pieces cancel completely between the
%%$\corr{E}{B}$ tails and the $\corr{E}{\chi}$ tails, another integrability miracle of the
%representation. Hence
%using $x_i\delta_{lj}+x_j\delta_{il}=\p_l(x_ix_j)$, $2x_l=\p_l r$, $4rx_l=\p_l r^2$.
%Valence $2$ admits one constant, and:
\end{itemize}
\paragraph{The complete solution:}
The Ward Identities associated with boost, translation, rotation and dilatation shape the two-point
functions of the null-reduction multiplet \eqref{eq:Sdata} completely, up to six constants, namely, 
$C_\psi$, $C_\chi$, $C_1$, $C_1'$, $C_B$, $C_E$. The collective correlation functions are given by
\begin{align*}
1.& ~\corr{\psi_i}{\psi_j}=C_\psi\,\delta_{ij}\,\tau^{-2\Delta},\\
2.&~\corr{\chi}{\chi}=C_\chi\,\tau^{-2\Delta}+C_\psi\,r\,\tp{2},\\
3.&~ \corr{\chi}{\psi_j}=\corr{\psi_j}{\chi}=-C_\psi\,x_j\,\tp{1},\\
4.&~\corr{B_{ij}}{\chi}=\corr{\chi}{B_{ij}}=0,\\
5.&~\corr{B_{ij}}{\psi_k}=\corr{\psi_k}{B_{ij}}
 =C_\psi\tp{1}(x_i\delta_{jk}-x_j\delta_{ik}),\hspace{-30mm}&&\\
6.&~\corr{B_{ij}}{B_{kl}}=C_B\,\tau^{-2\Delta}T_{ijkl}+C_\psi\,\tp{2}Q_{ijkl},&&\\
7.&~\corr{E_i}{\psi_j}=C_\psi\tp{2}\big(x_ix_j-\tfrac r2\delta_{ij}\big)+C_1\delta_{ij}\tau^{-2\Delta},\\
8.&~\corr{E_i}{\chi}=-(C_\chi{+}C_1)x_i\tp{1}-\tfrac{C_\psi}{2}x_i r\tp{3},\\
9.&~\corr{E_i}{B_{kl}}=\big[(C_1{-}C_B)\tp{1}-\tfrac{C_\psi}{2}r\tp{3}\big]
 \big(x_k\delta_{il}-x_l\delta_{ik}\big),\\
10.&~\corr{E_i}{E_j}=\text{eq.~\eqref{eq:EE}}.&&
\end{align*}
where $\corr{\psi_j}{E_i}, \corr{\chi}{E_i}, \corr{B_{kl}}{E_i}$ remains same as their counterpart with $C_1\to C_1'$. Nothing depends on this: only $C_1+C_1'$ enters \eqref{eq:EE}, and both constants vanish in the magnetic truncation. 
%The correlators are \emph{not} pure
%power laws: the multiplet is an indecomposable (Jordan-type) boost module, and the polynomial
%$x$-tails, growing in degree with the boost grade
%($\psi\psi$: degree $0$; $\chi\chi,B\psi,BB$: $\le2$; $E\chi,EB$: $\le3$; $EE$: $\le4$), are
%the position-space signature of its non-diagonalizable $M^{(0)}$ action.
\paragraph{Connection with Contraction Correlation functions:}
We will now see how two methods make a connection with each other. We have already seen at the level of equations of motion. Now, we will see it at the level of correlation functions. Let's divide it into three layers. They are given by:

\paragraph{(a) Truncation to the magnetic limit:}
To match it with the magnetic limit of the contraction method, we have to switch off the constants of the auxiliary fields, $C_\psi=C_\chi=C_1=C_1'=0$. It kills all correlators involving $\chi$ or $\psi$ and we are left with
\bea{}&&
\corr{B_{ij}}{B_{kl}}=C_B\tau^{-2\Delta}T,\qquad
\corr{E_i}{B_{kl}}=C_B\tp{1}(x_l\delta_{ik}-x_k\delta_{il}),\nonumber\\&&
\corr{E_i}{E_j}=C_E\delta_{ij}\tau^{-2\Delta}+C_B\tp{2}(r\delta_{ij}-x_ix_j).
\eea
These are exactly the magnetic limit \eqref{eq:magBB}--\eqref{eq:magEE}.

\paragraph{(b) On-shell $D=6$ truncation:}
The magnetic limit, as we know, enjoys the full GCA on the equations of motion at $D=6$. Similarly, this fact applies to the correlation functions. It gives, as we have seen, $C_B=0$,
\bea{}
\corr{E_i}{E_j}=C_E\,\delta_{ij}\,(t_1-t_2)^{-4},\qquad\text{all else }=0 .
\eea

\paragraph{(c) SCT (K) is unavailable for the full multiplet:}
When we had a look at the null-reduction theory, we understood that the action and equations are not SCT invariant. This was due to the presence of fields such as $(\chi, \psi_i)$. This makes the theory boost-conformal but not special-conformal. Similar thing is seen when we look at the correlation functions. One can see this when we impose \eqref{eq:K} on the pair $(\chi(1),\psi_j(2))$, it gives $\text{LHS}=-C_\psi(t_1{+}t_2)\delta_{kj}\tp{0}$,
$\text{RHS}=2t_1(-C_\psi\delta_{kj}\tp{0})$, residual $C_\psi(t_1-t_2)\delta_{kj}\tp{0}$. It collapses the tails. 

\subsection{Gauge-fixed propagator of the null-reduced theory}\label{gfp}
We start with \eqref{eq:LGKR} and add the null reduction of the covariant gauge-fixing term \cite{Townsend:1979hd, Siegel:1980jj, Son:2008ye}. It is given by 
\bea{}\mathcal L_{\rm gf}=-\frac1\xi(\partial^{\tilde\mu}B_{\tilde\mu\tilde\nu})(\partial_{\tilde\rho}B^{\tilde\rho\tilde\nu}).\eea 
With $\partial_u=0$ and $\eta^{ut}=1$, $V_{\tilde\nu}\equiv\partial^{\tilde\mu}B_{\tilde\mu\tilde\nu}$ has components $V_u=-\partial_i\psi_i$, $V_t=\partial_t\chi-\partial_iE_i$ and $V_j=\partial_t\psi_j+\partial_iB_{ij}$, and
\begin{equation}
\mathcal L_{\rm gf}=-\frac1\xi\big(2V_uV_t+V_jV_j\big)=\frac2\xi\,\partial_i\psi_i\big(\partial_t\chi-\partial_jE_j\big)-\frac1\xi\big(\partial_t\psi_j+\partial_iB_{ij}\big)^2 .
\end{equation}
The vector ghosts and the ghost-for-ghost scalars required by the reducible gauge symmetry decouple in the free abelian theory. We Fourier transform as 
\bea{}\Phi(t,\mathbf x)=\int\frac{d\omega\,d^{D-1}k}{(2\pi)^D}\,e^{-i\omega t+i\mathbf k\cdot\mathbf x}\,\Phi(\omega,\mathbf k).\eea 
Since $p^2=2p_up_t+\mathbf k^2$ and $p_u=0$, the Feynman-gauge ($\xi=1$) propagator is the parent one at $p_u=0$,
\begin{equation}\label{eq:prop}
\langle B_{\tilde\mu\tilde\nu}(\omega,\mathbf k)\,B_{\tilde\rho\tilde\sigma}(-\omega,-\mathbf k)\rangle=-\frac{i\,\big(\eta_{\tilde\mu\tilde\rho}\eta_{\tilde\nu\tilde\sigma}-\eta_{\tilde\mu\tilde\sigma}\eta_{\tilde\nu\tilde\rho}\big)}{2\,(\mathbf k^2-i\epsilon)} ,
\end{equation}
with non-vanishing components
\bea{}\label{eq:prop-comp}
\langle\chi\chi\rangle=\frac{i}{2\mathbf k^2},\qquad
\langle\psi_iE_j\rangle=\langle E_j\psi_i\rangle=-\frac{i\,\delta_{ij}}{2\mathbf k^2},\qquad
\langle B_{ij}B_{kl}\rangle=-\frac{i\,T_{ijkl}}{2\mathbf k^2},
\eea
while $\langle\psi\psi\rangle=\langle EE\rangle=\langle\chi\psi\rangle=\langle\chi E\rangle=\langle\chi B\rangle=\langle\psi B\rangle=\langle EB\rangle=0$. For every $\xi$ the denominators depend only on $\mathbf k$. The propagator is instantaneous, as for Galilean electrodynamics and Galilean Yang--Mills theory \cite{Chapman:2020vtn, Bagchi:2022eui}, and this is the precise sense in which the free theory does not determine $z$.

We can write the gauge-invariant correlators of $F_i$, $\Psi_{ij}$, $G_{ij}$ and $H_{ijk}$, and they are independent of $\xi$:
\begin{equation}\label{eq:fs-corr}
\begin{aligned}
\langle F_iF_j\rangle&=\frac{i\,k_ik_j}{2\mathbf k^2}, \qquad
\langle \Psi_{ij}G_{kl}\rangle=\frac{i\,Q_{ijkl}(\mathbf k)}{2\mathbf k^2}, \qquad
\langle G_{ij}G_{kl}\rangle=-\frac{i\,\omega^2\,T_{ijkl}}{2\mathbf k^2},\\
\langle F_iG_{jk}\rangle&=\frac{i\,\omega\,(k_j\delta_{ik}-k_k\delta_{ij})}{2\mathbf k^2}, \qquad
\langle G_{ij}H_{klm}\rangle=\frac{i\,\omega\,(k_kT_{ijlm}+k_lT_{ijmk}+k_mT_{ijkl})}{2\mathbf k^2},\\
\langle H_{ijk}H_{lmn}\rangle&=-\frac{i}{2\mathbf k^2}\sum_{(abc)}\sum_{(def)}k_a k_d\,T_{bcef},\qquad
\langle \Psi\Psi\rangle=\langle F\Psi\rangle=\langle FH\rangle=\langle\Psi H\rangle=0,
\end{aligned}
\end{equation}
where $(abc)$ and $(def)$ run over the cyclic permutations of $(ijk)$ and $(lmn)$, and all correlators are $\langle X(\omega,\mathbf k)\,Y(-\omega,-\mathbf k)\rangle$. Here $Q_{ijkl}(\mathbf k)$ is (6.15) with $x\to k$. Because $\omega$ enters only polynomially, position-space correlators are derivatives of
\bea{}
\delta(\tau)\,\mathcal G(\mathbf x),\qquad \mathcal G(\mathbf x)=\int\frac{d^{D-1}k}{(2\pi)^{D-1}}\frac{e^{i\mathbf k\cdot\mathbf x}}{\mathbf k^2}=\frac{\Gamma\!\big(\frac{D-3}{2}\big)}{4\pi^{\frac{D-1}{2}}\,|\mathbf x|^{D-3}} ;
\eea
for example $\langle B_{ij}(t,\mathbf x)B_{kl}(0)\rangle=-\frac{i}{2}\,T_{ijkl}\,\delta(t)\,\mathcal G(\mathbf x)$. Since $\delta(\tau)$ carries weight $z$, these correlators have total weight $z+D-3$, which matches $\Delta_\Phi+\Delta_{\Phi'}$ from \eqref{eq:GKR-dims} for \emph{every} $z$. Because $\tau\,\delta(\tau)=0$, the boost identity \eqref{eq:B} reduces to $\langle S_l[\Phi_A]\Phi_B\rangle+\langle\Phi_AS_l[\Phi_B]\rangle=0$. The propagator (\ref{eq:prop-comp}) satisfies this for all pairs, for instance $\langle\delta_{il}\psi_jE_k-\delta_{jl}\psi_iE_k\rangle+\langle B_{ij}B_{kl}\rangle=0$. Since $t_1=t_2$ on the support, the special conformal identity \eqref{eq:K} is then satisfied as well. The free theory thus realises a distributional solution of the Ward identities of that lies outside the power-law ansatz \eqref{eq:dil}.
\section{Summary and Future Directions}
\label{sec:summary}
In this paper, we construct and understand the Galilean limit of Kalb-Ramond theory. We first looked into the relativistic conformal transformations and also the Kalb-Ramond field theory in detail. We found that the theory is conformally invariant in $D=6$. The relativistic theory only has finite symmetry in $D=6$. Next, we see the non-relativistic limit of conformal symmetry, known as the Galilean conformal algebra. It was found by taking the Inönü-Wigner contraction. We talk about two methods to take the Galilean limit on Kalb-Ramond theory. First, was found by taking the scaling of space-time coordinates as well as the two-forms. Second, we used the null-reduction method. The representation theory was discussed in detailed for both the methods. In the first case, we did the splits of two-form into ${E_i, B_{ij}}$ and it yields two limits. These limits are known as Electric and Magnetic limits. In both, the equations of motion carry the full infinite-dimensional GCA precisely at $D = 6$.

For the second case, we have performed a null reduction from $D+1$ to $D$ dimensions, resulting in a Lagrangian that contains additional bosonic fields: $\chi$ and $\psi_i$. The Lagrangian and equations are invariant under all global generators in the zero-momentum sector: $n=-1,0$ for all $D$. However, they are not invariant under special conformal generators in the GCA or higher modes. This is different to Galilean electrodynamics, and we trace this back to the presence of the vector $\psi_i$ at the bottom of the boost chain. The symmetries induced by the parent conformal group form a Schr\"odinger algebra extended by an additional dilatation generator, hence $z$ cannot be fixed. In the case that the parent theory is the critical six-dimensional theory, the five-dimensional reduced action is invariant under the Schr\"odinger special conformal transformation with $z=2$. The non-zero null momentum sectors are free Schr\"odinger fields. Therefore, both methods highlight a special critical dimension. The contraction method yields the critical dimension due to the presence of GCA in the contraction at $D=6$. The null reduction method results in a special dimension due to the induced Schr\"odinger symmetry in the null reduction at $D+1=6$. In terms of correlation functions, the Ward identities imply that the contracted correlation functions are fixed to a single ultra-local power law under the special conformal generator. The correlation functions in the null reduced module contain six constants. These constants create polynomial tails down the nilpotent boost chain: $E_i\to\{B_{ij},\chi\}\to\psi_i\to0$. However, in the free null reduced theory there is no frequency-dependent propagator, and hence the correlation functions have support at $\tau=0$.

We will give some future directions for this project.
\paragraph{Canonical analysis and the central extension:} We have realized the infinite symmetry enhancement in this paper. The next course is the Hamiltonian analysis of the null-reduced action \eqref{eq:LGKR}. In this, we want to study the reducible gauge symmetry \cite{Henneaux:1992ig}, the Dirac brackets, the Noether charges, and their algebra \cite{Bagchi:2022eui,Banerjee:2019axy}. This will help to determine a few things:
\begin{enumerate}
    \item Fixation of the physical degree-of-freedom count.
    \item Understand whether the higher modes are recovered as canonical charges with a central
term \cite{Banerjee:2019axy,Bagchi:2019clu}.
\end{enumerate}
This is the most immediate technical task to be done.
\paragraph{The general $p$-form and the pattern of the obstruction:}
The method we have talked in this paper can be extended to a $p$-form gauge field $A_{\mu_1\cdots\mu_p}$
with $(p{+}1)$-form field strength. The relativistic theory is conformal only in dimension $D=2(p+1)$ \cite{Jackiw:2011vz,ElShowk:2011gz,Nakayama:2013is}. This is known as the critical dimension. If we take $p=1$ or $D=4$, it is the Maxwell case \cite{Bagchi:2014ysa,Festuccia:2016caf,Mehra:2021sfx}. Similarly, if we take  $p=2$ ($D=6$), it is the Kalb--Ramond case \cite{Kalb:1974yc}. These are the first two members of this series.
The most important feature that is
absent in $p=1$ is that the action loses the special-conformal generators and the higher modes \cite{Bagchi:2022eui}. We want to understand 
if this feature is true for $p\ge 2$, and how its severity scales with $p$ \cite{Henneaux:2021yzg,Bergshoeff:2020xhv}? This is a very straightforward statement which is internal to the field theory, and independent of any external input.
%The construction extends verbatim to a $p$-form gauge field $A_{\mu_1\cdots\mu_p}$
%with $(p{+}1)$-form field strength, for which the relativistic theory is
%%conformal in the critical dimension $D=2(p+1)$; the Maxwell case $p=1$ ($D=4$)
%and the present Kalb--Ramond case $p=2$ ($D=6$) are the first two members. 
%The distinctive feature found here---that the free action loses the special-conformal (or higher modes)
%generators through a \emph{dynamical tensor component sitting under $\dd_t$ with
%%non-vanishing boost-spin} (the $\Psi_{lj}\psi_j$ obstruction), a mechanism
%absent for $p=1$---raises a sharp, self-contained question: 
%
%does the action-level breakdown of the special-conformal (and higher Witt) modes switch
%on precisely for $p\ge 2$, and how does its severity scale with $p$? This is a
%clean representation-theoretic statement, internal to the field theory, and
%independent of any external input.

\paragraph{DLCQ of six-dimensional theories:} 
We compactify the null direction as $u\sim u+2\pi R$ so it then quantises the null momentum, $m=N/R$. The reduced theories of Sec\eqref{sec:nullreduction} then become the sectors of the discrete light-cone quantisation (DLCQ). For $D+1=6$, one finds that the sectors $N\neq0$ are precisely those applied to light-cone quantisation of the $(2,0)$ superconformal theory \cite{Aharony:1997an,Lambert:2011gb}. 
Setting $N$ reduces this to a quantum-mechanical description with the moduli space of instantons as its phase space. For the free abelian tensor multiplet, the two-form contributes only those fields of Sec\eqref{snznm} that come from the Schr\"odinger action, specifically only their self-dual part. The $N=0$ sector studied in Sec\eqref{sec:nullreduction} is the zero mode. In DLCQ, this mode is known to be constrained rather than dynamical, consistent with the frequency-independent propagator of Sec\eqref{sec:nullreduction}, and its Schr\"odinger symmetry at $D+1=6$ is the remnant of six-dimensional conformal invariance. There has recently been study of null reductions of the M5-brane world volume theory \cite{Lambert:2020zdc, Lambert:2019jwi}, obtaining an enhanced spacetime symmetry in five non-Lorentzian dimensions. It would be interesting to embed the free two-form results obtained here into those constructions and into the DLCQ of little string theories \cite{Aharony:1997th}. Those six-dimensional theories are not conformal, and the inherited symmetry is expected to reduce to the Bargmann subalgebra.

\paragraph{Coupling to Tensionless strings and the two dualities:}
We have seen that the background field to which the string couples is the Kalb-Ramond field \cite{Kalb:1974yc,Oling:2022fft,Harmark:2017rpg}, whose term is given by $\int_\Sigma B_{\mu\nu}\,dx^\mu\wedge dx^\nu$. When one considers the tensionless regime \cite{Bagchi:2013bga,Bagchi:2015nca}, such backgrounds give a nontrivial contribution. It deforms the compactified spectrum and partially restores $O(d,d)$ duality \cite{Banerjee:2023ekd}. The worldsheet symmetry is, in this case, given by $\mathrm{BMS}_3$ \cite{Banerjee:2024fbi}. The $2$-form constructed in this paper is the field whose zero mode is that background. Therefore, one has two questions:
\begin{enumerate}
\item Does the multiplet $\{\chi,\psi_i, E_i, B_{ij}\}$ arises as the vertex-operator data deforming the worldsheet coupling? Do we get the equality between worldsheet $\mathrm{BMS}_3$ and the target-space infinite GCA \cite{Bagchi:2009my,Bagchi:2016geg,Bagchi:2019xfx}?
\item Does the electric-magnetic duality match with the target-space face
of the worldsheet T-duality \cite{Banerjee:2024fbi,Banerjee:2023ekd}?
\end{enumerate}
These are the most speculative directions and worth pursuing once the free-field structure is understood well.
\subsection*{Acknowledgements}
It is a pleasure to thank colleagues at Christ University for discussions.
\bibliographystyle{unsrt}
\bibliography{Untitled}
%\begin{thebibliography}{9}
%\bibitem{KalbRamond} M.~Kalb and P.~Ramond,
%``Classical direct interstring action,''
%Phys.\ Rev.\ D \textbf{9} (1974) 2273.
%\bibitem{JackiwPi} R.~Jackiw and S.-Y.~Pi,
%``Tutorial on scale and conformal symmetries in diverse dimensions,''
%J.\ Phys.\ A \textbf{44} (2011) 223001, arXiv:1101.4886.
%\bibitem{ElShowk} S.~El-Showk, Y.~Nakayama and S.~Rychkov,
%``What Maxwell theory in $D\neq4$ teaches us about scale and conformal invariance,''
%Nucl.\ Phys.\ B \textbf{848} (2011) 578, arXiv:1101.5385.
%\bibitem{BBM} A.~Bagchi, R.~Basu and A.~Mehra,
%``Galilean conformal electrodynamics,''
%JHEP \textbf{1411} (2014) 061, arXiv:1408.0810.
%\bibitem{LBLL} M.~Le Bellac and J.-M.~L\'evy-Leblond,
%``Galilean electromagnetism,''
%Nuovo Cimento \textbf{14B} (1973) 217.
%
%\end{thebibliography}

\end{document}